\documentclass[10pt, Conference, letterpaper]{IEEEtran}
\IEEEoverridecommandlockouts
\usepackage{cite}
\usepackage{lipsum} % For dummy text
\usepackage{booktabs}
\usepackage[T1]{fontenc}
\usepackage{pifont}
\usepackage{aecompl}
\usepackage[numbers,sort&compress]{natbib}
\usepackage{amsmath,amssymb,amsfonts}
\usepackage{amsthm}
\usepackage{float} %提供float浮动环境
\usepackage{booktabs} %提供命令\toprule、\midrule、\bottomrule
\makeatletter
\renewcommand{\maketag@@@}[1]{\hbox{\m@th\normalsize\normalfont#1}}%
\makeatother
\usepackage{graphicx}
\usepackage{verbatim}
\usepackage{subcaption}
\usepackage{textcomp}
\usepackage{xcolor}
\usepackage{stfloats}
\usepackage{tabularx}
\usepackage{array}
\usepackage{makecell}
\usepackage{url}
\makeatother
\usepackage{textcomp}
\usepackage[marginal]{footmisc}
\usepackage{geometry}
\usepackage[hidelinks]{hyperref}
\usepackage{setspace}
\usepackage{caption}
\allowdisplaybreaks[4]
\usepackage{algpseudocode}
\usepackage[ruled,linesnumbered]{algorithm2e}                %算法排版样式1

\def\BibTeX{{\rm B\kern-.05em{\sc i\kern-.025em b}\kern-.08em
    T\kern-.1667em\lower.7ex\hbox{E}\kern-.125emX}}

\begin{document}
\title{DMSA: A Decentralized Microservice Architecture for Edge Networks}
% \title{DMSA: A Decentralized Microservice Architecture for Accurate Awareness, Low Monitoring Overhead, and Dynamic Scheduling in Edge Networks}
\author{Yuang Chen, \IEEEmembership{Graduate Student Member, IEEE}, Chengdi Lu, Yongsheng Huang, Chang Wu, Fengqian Guo, Hancheng Lu, \IEEEmembership{Senior Member, IEEE}, and Chang Wen Chen, \IEEEmembership{Fellow, IEEE}
% \thanks{\setlength{\baselineskip}{1.5\baselineskip} Yuang Chen, Chengdi Lu, Yongsheng Huang, Chang Wu, Fengqian Guo, and Hancheng Lu are with the University of Science and Technology of China, Hefei 230027, China. (email: yuangchen21@mail.ustc.edu.cn; lcd1999@mail.ustc.edu.cn; ysh6@mail.ustc.edu.cn; changwu@mail.ustc.edu.cn; fqguo@ustc.edu.cn; hclu@ustc.edu.cn). Hancheng Lu is also with the Institute of Artificial Intelligence, Hefei Comprehensive National Science Center, Hefei 230088. Yuang Chen is also with the Department of Computing, The Hong Kong Polytechnic University, China (email: yuang.chen@polyu.edu.hk). Chang Wen Chen is with the Department of Computing, The Hong Kong Polytechnic University, China. (email: changwen.chen@polyu.edu.hk)}
\thanks{Yuang Chen, Chengdi Lu, Yongsheng Huang, Chang Wu, Fengqian Guo, and Hancheng Lu are associated with the University of Science and Technology of China in Hefei 230027, China. You can reach them via email at yuangchen21@mail.ustc.edu.cn, lcd1999@mail.ustc.edu.cn, ysh6@mail.ustc.edu.cn, changwu@mail.ustc.edu.cn, fqguo@ustc.edu.cn, and hclu@ustc.edu.cn respectively. Prof. Hancheng Lu is also affiliated with the Institute of Artificial Intelligence, Hefei Comprehensive National Science Center, Hefei 230088. Prof. Chang Wen Chen holds the position of Chair Professor at the Department of Computing at The Hong Kong Polytechnic University in China and can be reached at changwen.chen@polyu.edu.hk.}
}
\maketitle
\pagestyle{empty} %no page number for the second and the later
\thispagestyle{empty}% no page number for the first page
\begin{abstract}
The dispersed node locations and complex topologies of edge networks, combined with intricate dynamic microservice dependencies, render traditional centralized microservice architectures (MSAs) unsuitable. In this paper, we propose a decentralized microservice architecture (DMSA), which delegates scheduling functions from the control plane to edge nodes. DMSA redesigns and implements three core modules of microservice discovery, monitoring, and scheduling for edge networks to achieve precise awareness of instance deployments, low monitoring overhead and measurement errors, and accurate dynamic scheduling, respectively. Particularly, DMSA has customized a microservice scheduling scheme that leverages multi-port listening and zero-copy forwarding to guarantee high data forwarding efficiency. Moreover, a dynamic weighted multi-level load balancing algorithm is proposed to adjust scheduling dynamically with consideration of reliability, priority, and response delay. Finally, we have implemented a physical verification platform for DMSA. Extensive empirical results demonstrate that compared to state-of-the-art and traditional scheduling schemes, DMSA effectively counteracts link failures and network fluctuations, improving the service response delay and execution success rate by approximately $60\% \sim 75\%$ and $10\%\sim15\%$, respectively.
\end{abstract}
\vspace{-0.5em}

\vspace{-0.5em}

\section{Introduction}
\par With the burgeoning advancement of cloud-native techniques such as Spring Native, Swarm, and Docker, alongside the exponential growth of digital content and services, microservice architectures (MSAs) have emerged as formidable enablers for flexible service deployment in distributed scenarios \cite{8990350, vstefanivc2019switch, hannousse2021securing, pahl2016microservices, 8951173,zhang2024}. For instance, Microsoft Exchange leverages hundreds of microservices to facilitate efficient email delivery \cite{zeng2023traceark}. Unlike traditional service-oriented architectures (SOA) that rely on enterprise service buses (ESB), MSAs decompose large-scale monolithic applications into massive re-deployable, single-purpose, and lightweight microservices \cite{waseem2020systematic, 9615028, gu2022layer, zeng2023layered}. These microservices characterized by intricate dynamic dependencies, communicate and interact to collaboratively handle user requests, thereby facilitating flexible development and deployment, and enhancing system scalability and reliability \cite{9615028, gu2022layer, zeng2023layered,9057418}. In addition, as mobile application providers and terminal devices demand higher responsiveness and real-time experiences, traditional cloud computing can no longer fulfill the low end-to-end service invoking delay requirements \cite{8391395}. As an advanced post-cloud paradigm, edge computing enhances real-time performance and reduces data traffic by positioning computation and memory resources closer to users \cite{10430407,10460318,10529607}.

\par Customizing effective microservice scheduling for edge networks holds the promise to significantly improve communication latency and service response speed, while ensuring stable operation. However, integrating MSAs into edge networks also presents substantial challenges. Unlike cloud computing, edge networks have limited computing and memory resources, dispersed node locations, intricate network topologies, high user mobility, and device heterogeneity \cite{zhang2023resource}. It is also due to the multi-access and distributed nature of edge computing, to fully utilize the resources, edge computing necessitates reallocating hardware and software to serve multiple users on the same hardware. This is typically achieved by leveraging virtualization techniques. However, traditional virtual machine (VM)-based virtualization techniques are impractical for edge networks due to slow startup times, high resource overhead, and complex deployment and migration processes \cite{xu2013managing, 8967018}. Fortunately, microservices are usually deployed based on lightweight virtualization techniques like \emph{container} \cite{10098822}. It provides faster deployment and startup speeds, lower resource consumption, and stronger compatibility, and can be flexibly deployed on resource-constrained, platform-heterogeneous edge or fog devices \cite{pallewatta2022qos}. However, implementing flexible microservice scheduling in edge networks generally involves complex topology forwarding, and the limited and unstable link bandwidth can lead to significant performance variations across nodes and instances. Therefore, edge networks require more comprehensive and efficient microservice schemes to guarantee the quality-of-service (QoS) of applications \cite{10379832, zeng2023layered}.

\begin{figure}[t]
\vspace{-0.3em}
\centering
\includegraphics[scale=0.33]{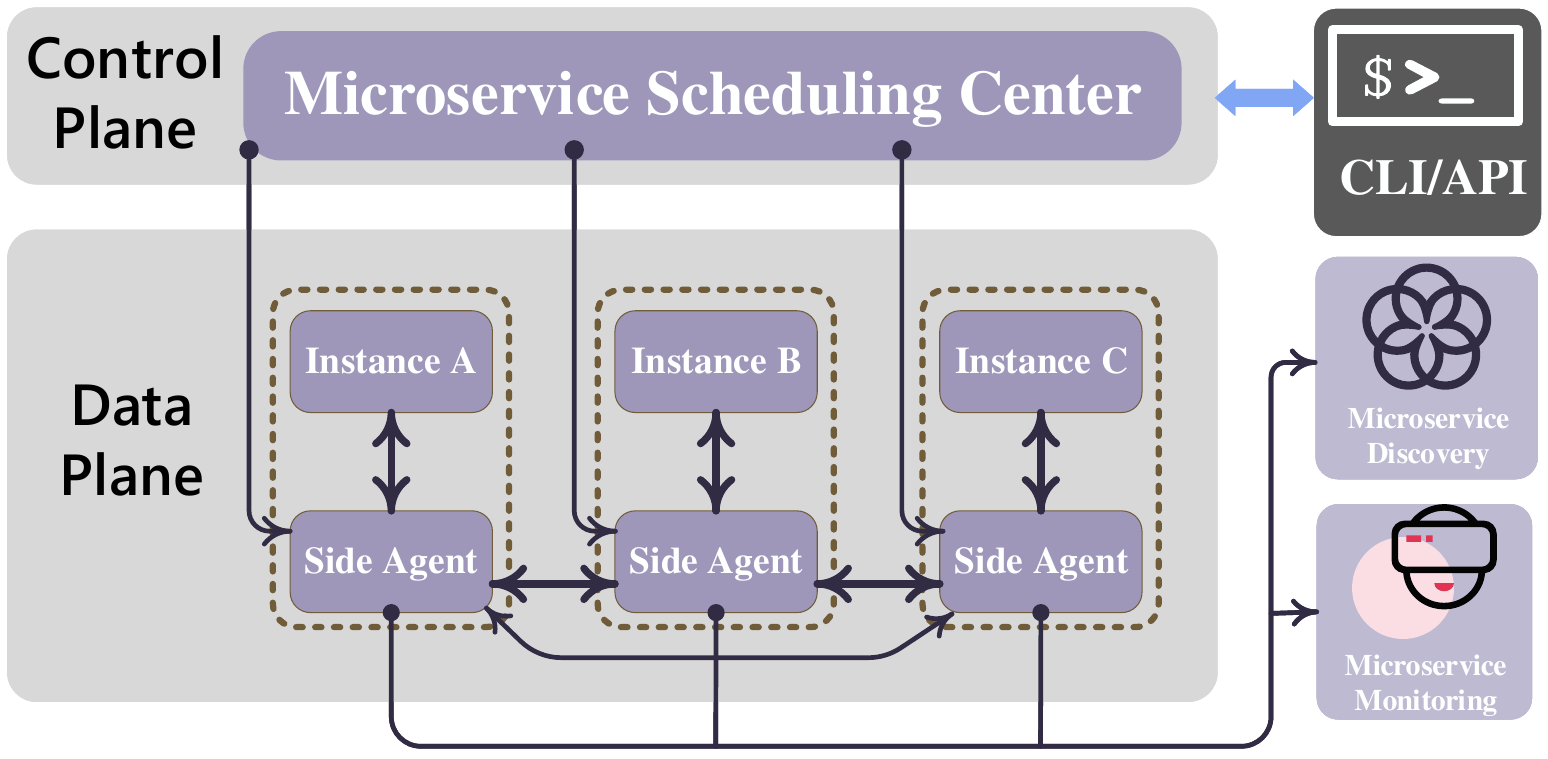}
\vspace{-0.5em}
\caption{\small Traditional centralized microservice architecture.}
\label{fig1}
\vspace{-1em}
\end{figure}

\par As illustrated in Fig. \ref{fig1}, the vast majority of current studies rely on traditional centralized control-based MSAs \cite{vstefanivc2019switch,waseem2020systematic,9615028,pahl2016microservices,hannousse2021securing,zeng2023layered,9057418,10379832}. The control plane collects instance information from the data plane and computes scheduling strategies. The side agents in the data plane then process and forward instance requests according to the strategies issued by the scheduling center. This setup makes the functions of microservice discovery, load balancing, monitoring and scheduling heavily dependent on the scheduling center \cite{9615028,9869329}. Take \emph{Nautilus} as an example \cite{9615028,10592806,9820678}, which considers the communication overhead and resource contention between microservices and external co-located tasks. Its central scheduler continuously monitors the CPU, memory, and network load of each node, thereby effectively realizing the deployment of microservice-based user-facing services in the cloud-edge continuum\cite{9460542,9615028,10592806,9820678}. Nonetheless, in edge networks with unstable link states, centralized control is prone to single points of failure \cite{10229031}. Any interruption of connection to the control plane can result in the inability of side agents to receive timely updates, potentially leading to complete system paralysis and severely impacting the scheduling capability of edge nodes. In addition, the dispersed nature of edge devices means some nodes may be distant from the control center, leading to high control overhead. Consequently, it is imperative to meticulously customize the MSA for edge networks and to redesign and implement the relevant core functional modules.

\begin{figure}[t]
\vspace{-0.3em}
\centering
\includegraphics[scale=0.25]{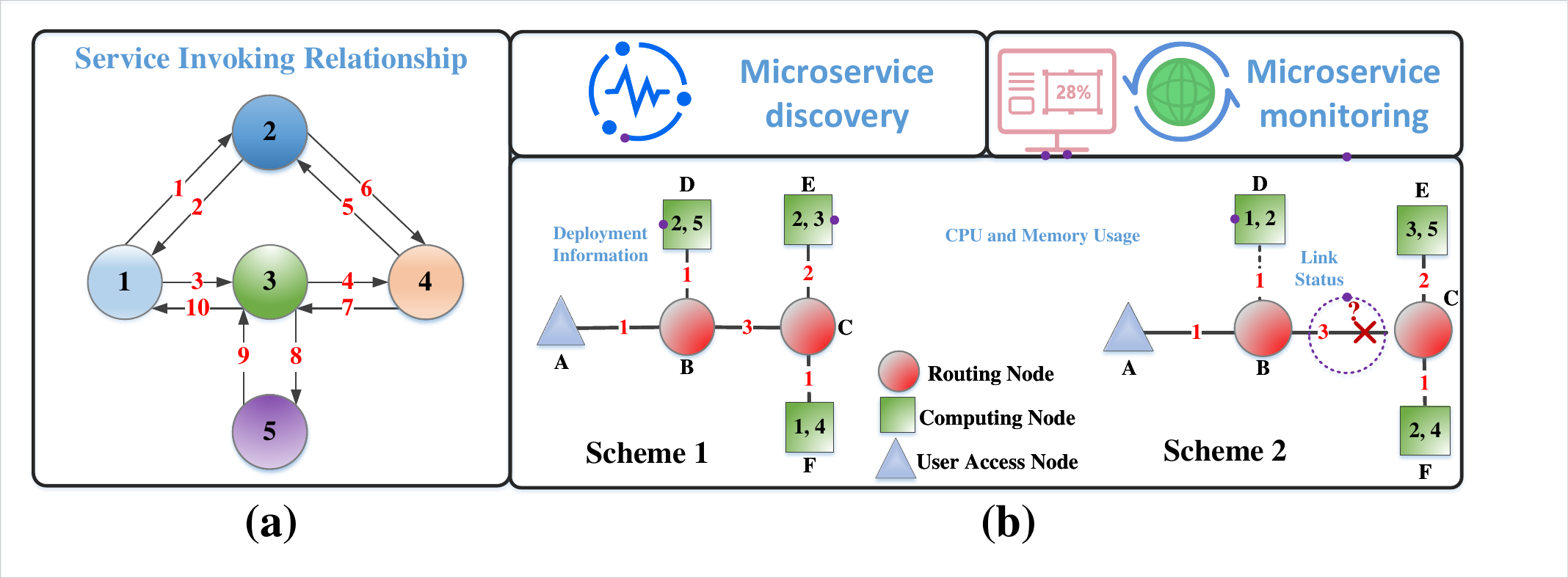}
\vspace{-0.3em}
\caption{\small The performance impact of microservice discovery, monitoring, and scheduling. (a) Microservice Invocations. (b) Two different schemes for the depicted microservice invocation.}
\label{fig2}
\vspace{-0.5em}
\end{figure}

\par As illustrated in Fig. \ref{fig2}, we analyze the performance impact of microservice discovery, monitoring, and scheduling. Fig. \ref{fig2} (a) shows the microservice invocations for a specific service, with node labels indicating microservice types and arrow labels denoting the order in which microservices are invoked. Fig. \ref{fig2} (b) presents two deployment schemes for this specific microservice invocation, with the node labels identifying the deployed microservice instances and the connecting lines indicating the communication delay (in \emph{ms}). As depicted in Fig. \ref{fig2} (b), before scheduling, microservice discovery senses and logs the deployment of instances across nodes. During operation, microservice monitoring collects load messages, e.g., CPU and memory usage from instances and nodes in real-time, and monitors link status such as available bandwidth, response delay, etc \cite{9615028,8685146}. Compared to Scheme 1, Scheme 2 significantly reduces communication delays by deploying frequently interacting instances on the same or nearby computing nodes. Thus, the communication delay from user-access node \textbf{A} to the response is $22$ \emph{ms}, while in Scheme 1 it is $46$ \emph{ms}. As a result, microservice scheduling is essential for performance improvement. For example, Scheme 2 deploys instances of microservice 2 on both compute nodes \textbf{D} and \textbf{F} to enable microservices 1 and 4 to make local invocations, thus avoiding cross-node communication and reducing delay and overhead. Consequently, effective microservice scheduling must account for inter-node communication overhead, load balancing, and design MSAs tailored to edge network characteristics to improve response speed and service stability.

\par To effectively overcome the aforementioned challenges, we have customized an innovative decentralized microservice architecture (DMSA) for edge networks. For the first time, the proposed DMSA sinks the microservice scheduling functions onto the edge nodes. It re-designs and implements the three core functional modules of microservice discovery, monitoring, and scheduling, thereby significantly enhancing the response speed of microservice scheduling and its robustness against unexpected changes in edge network topologies. The primary contributions of this paper are summarized as follows:

\begin{itemize}
  \item We have proposed an innovative microservice architecture called DMSA, which delegates the scheduling function from the control plane to microservice agent (MA)-equipped edge nodes. DMSA redesigns and implements the three core functions of microservice discovery, monitoring, and scheduling modules. These MAs are independent of each other to avoid the impact of the single points of failure on other nodes.

  \item In DMSA, the discovery module achieves precise sensing of instance deployments and complex dependencies through efficient interactions with microservice configuration files and instance update messages. For the monitoring module, the CPU and memory usage of instances is directly obtained through Docker container-provided interfaces to monitor them continuously. In particular, a novel publisher-subscriber model-based status information synchronization (PSMS) mechanism is tailored for DMSA to facilitate low monitoring overhead. Moreover, a combined active and passive (CAPA) mechanism is developed to mitigate bandwidth measurement errors.

  \item Moreover, DMSA has meticulously customized the microservice scheduling scheme, where multi-port listening and zero-copy forwarding techniques are leveraged to guarantee the portability and high data forwarding efficiency of DMSA. Particularly, a dynamic weighted multilevel load balancing (DWMLB) algorithm is proposed to accurately and dynamically adjust scheduling strategies with the consideration of the microservice's reliability, priority, and response delay.
  
  \item Finally, we have constructed a physical verification platform for DMSA to evaluate its performance in complex edge network topologies. Extensive empirical results demonstrate that compared to the state-of-the-art scheduling scheme Nautilus \cite{9615028, 9460542}, as well as the generalized scheduling schemes, i.e., Least Connects \cite{yang2024reducing} and Round Robin \cite{nasser2016analisis}, DMSA effectively withstands link failures and network fluctuations, significantly improving the service response delay and execution success rate by about $60\% \sim 75\%$ and $10\% \sim 15\%$, respectively.
\end{itemize}

\par The remainder of this paper is organized as follows. Sec. II introduces the design of DMSA. Sec. III explicates the implementation of DMSA's key modules. In Sec. IV, we build the physical verification platform to test and analyze the performance of DMSA in practical edge networks with complex node topology. Finally, Sec. V concludes the paper and explores future directions.

\vspace{-0.6em}

\section{Design OF THE DECENTRALIZED MICROSERVICE ARCHITECTURE}
\par
\begin{figure}[h]
\centering
\includegraphics[scale=0.30]{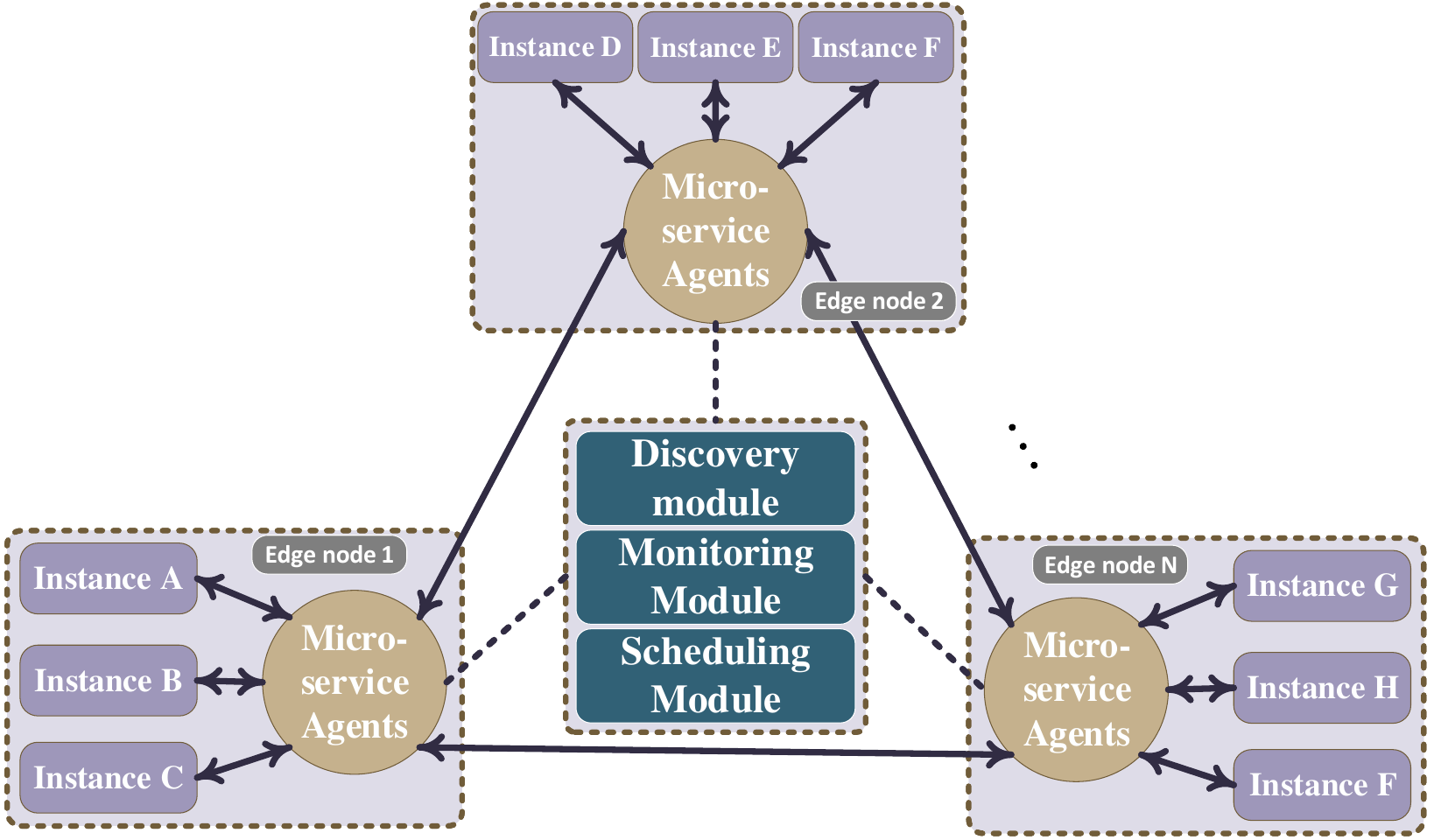}
\vspace{-0.3em}
\caption{\small The decentralized microservice architecture.}
\label{fig3}
\vspace{-0.5em}
\end{figure}

 Fig. \ref{fig3} illustrates the proposed DMSA that eliminates the centralized control plane and sinks microservice scheduling to each MA-equipped edge node. The communication among microservice instances is completed by the MA of the host node. The MA dynamically adjusts scheduling strategies through inter-agent information exchanges and network status awareness, and forwards instance requests based on scheduling strategies. These MAs operate independently and autonomously at the transport layer to avoid affecting other nodes due to single points of failure. To facilitate microservice monitoring and scheduling, the MA provides both internal and external bidirectional proxy services, which internally forward the service requests of the local instances to corresponding target microservice instances and externally forward access requests to instances on this node. In this paper, the MA exploits the non-invasive framework to guarantee flexibility and scalability\cite{shen2023network,10428037}\footnote{MSAs typically use either intrusive or non-intrusive frameworks \cite{shen2023network,10428037}. Intrusive frameworks involve modifying MSA's original code, which not only increases extra workload for developers but also contradicts the vision of a high degree of decoupling among microservice modules. Conversely, non-intrusive frameworks manage microservices by capturing external interfaces without altering the original code, requiring only simple configuration \cite{shen2023network,10428037}.}, which only modifies the header of each request or response without inspecting its content. During operation, the MA uses multiple internal and external ports based on configuration information, and local microservices specify the next microservice type to be invoked by requesting different proxy ports.

\vspace{-0.5em}

\par
\begin{figure}[h]
\vspace{-0.3em}
\centering
\includegraphics[scale=0.45]{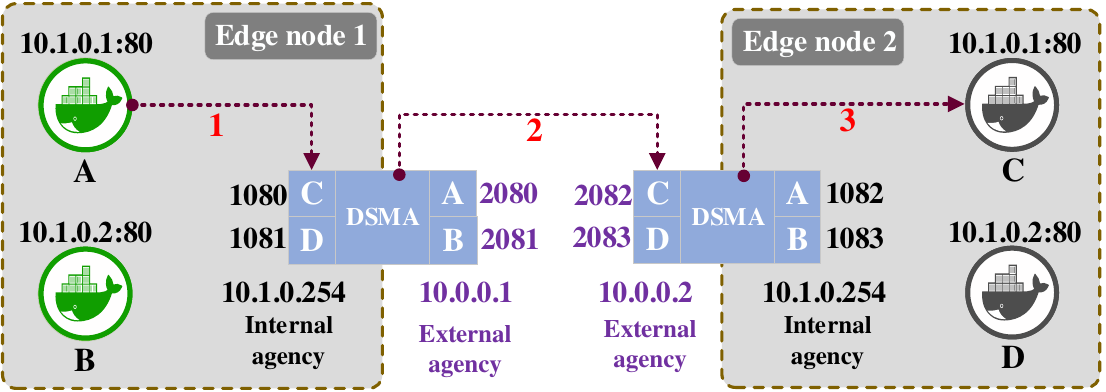}
\vspace{-0.2em}
\caption{\small The request forwarding process of DMSA.}
\label{fig4}
\vspace{-1em}
\end{figure}

 Fig. \ref{fig4} illustrates the specific process of MAs forwarding microservice requests. Microservice \textbf{A} on edge node \textbf{1} accesses port \textbf{1080} of the local MA to notify DMSA to access Microservice \textbf{C}. Upon receiving the notification, the MA on edge node \textbf{1} forwards it to port \textbf{2082} on edge node \textbf{2} according to the scheduling strategy. The MA on edge node \textbf{2} determines the target microservice \textbf{C} based on the port number and forwards the request to the instance of Microservice \textbf{C} on this node.

\section{Implementation of Core Modules for DMSA}
\par
\begin{figure}[H]
\vspace{-0.3em}
\centering
\includegraphics[scale=0.5]{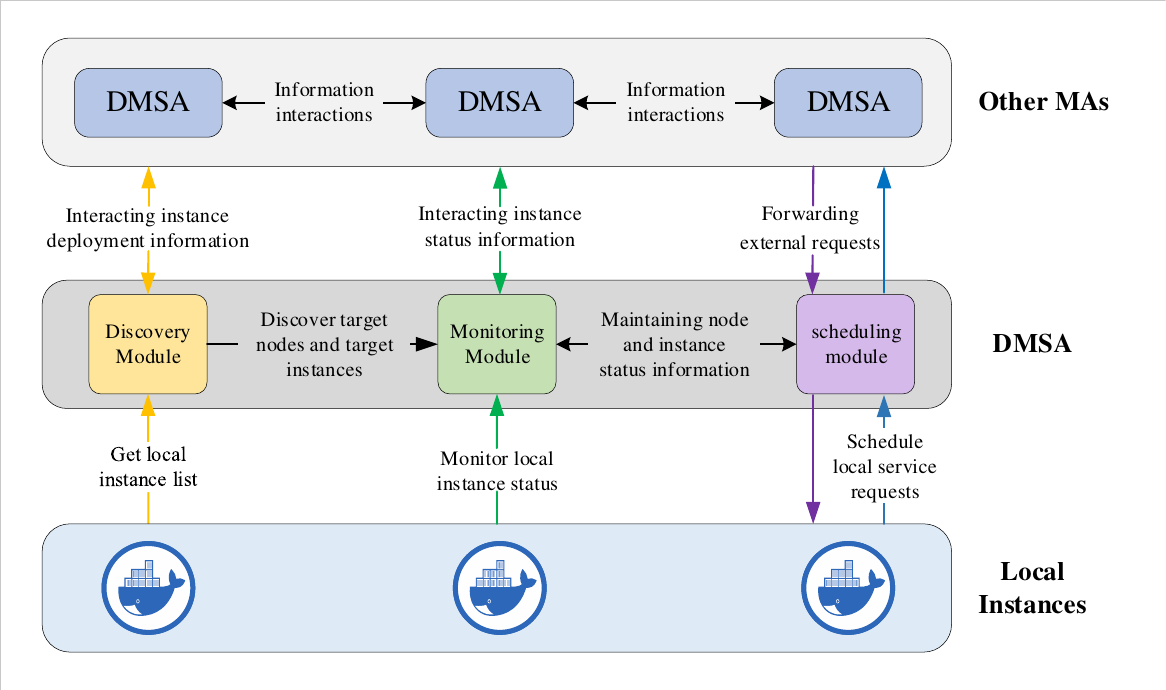}
\vspace{-0.2em}
\caption{\small The relationships among DMSA's key functional modules.}
\label{fig5}
\vspace{-1.5em}
\end{figure}

 Traditional centralized MSAs rely heavily on the control center for service discovery and load balancing. This is prone to single points of failure that can cripple the system \cite{vstefanivc2019switch,waseem2020systematic,9615028,pahl2016microservices,hannousse2021securing,zeng2023layered,9057418,10379832}. To this end, DMSA has made architectural innovations by redesigning and implementing microservice discovery, monitoring, and scheduling modules. Fig. \ref{fig5} illustrates the relationship between these core functional modules. To facilitate the subsequent description, we give the following definitions:

\begin{itemize}
  \item \textbf{Local Microservice:} The microservice deployed on the current MA's located edge node.

  \item \textbf{Target Microservices:} Microservices that have dependencies with the local microservice.

  \item \textbf{Target node:} The node where the target microservice instance is deployed.
\end{itemize}

\vspace{-0.5em}

\subsection{Implementation of Microservice Discovery Module}
\par
\begin{figure}[H]
\vspace{-0.3em}
\centering
\includegraphics[scale=0.4]{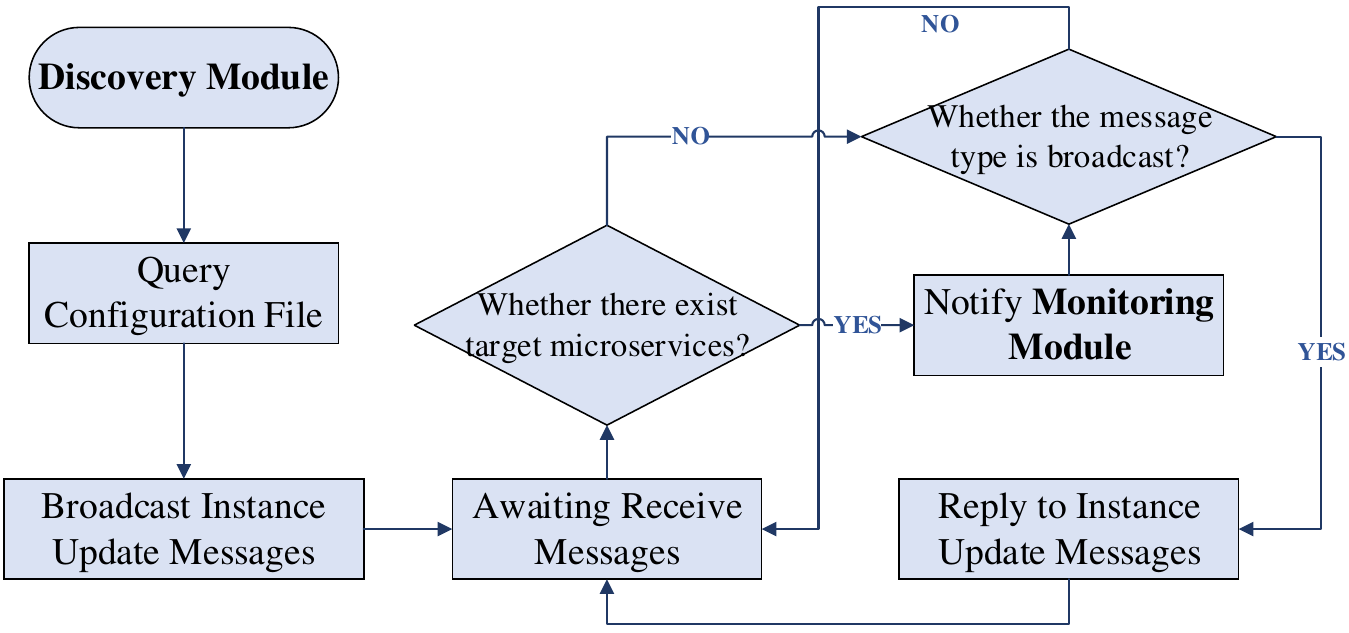}
\vspace{-0.3em}
\caption{\small The workflow diagram of the microservice discovery module.}
\label{fig6}
\vspace{-1.2em}
\end{figure}

 The prerequisite for realizing efficient microservice scheduling is to effectively discover microservices. This requires accurate identification of instance deployments as well as network and node status, as shown in Fig. \ref{fig5}. In this paper, we achieve this through the interaction of configuration files and instance update messages, as detailed in Fig. \ref{fig6}. Each MA initially senses the deployment and dependency information of microservices on its host node by utilizing configuration files, which comprise two parts: the local microservices list and the target microservices list. The local microservices list includes the ID type, name, container IP and port, and external port of each local instance, as shown in Table \ref{tab1}. The target microservices list specifies the type ID and internal agent port of microservices that local microservices require to invoke. Local microservices can invoke the required microservices by accessing the corresponding proxy ports. Notably, DMSA does not internally proxy all types of microservices because not all microservices have dependencies on each other.

\vspace{-1em}

\begin{table}[h]
\setlength{\belowcaptionskip}{-2bp}
\centering
\captionsetup{font=footnotesize} % further reduce font size of caption
\caption{Description of Main Fields in the Configuration File}
\label{tab1}
\renewcommand{\arraystretch}{0.8} % further reduce row height
\setlength{\tabcolsep}{3pt} % further reduce column spacing
\begin{tabular}{ll}
\toprule
Field     & Description                             \\
\midrule
TypeID    & Microservice Type ID                    \\
Name      & Microservice Instance Name              \\
Address   & IP Address of the Instance's Container  \\
LocalPort & Port Used by the Instance for Service   \\
ProxyPort & Port Used by DMSA for Proxy Service     \\
\bottomrule
\end{tabular}
\vspace{-1.0em}
\end{table}

\par
\begin{figure}[H]
    \centering
    \begin{subfigure}{0.218\textwidth}
        \centering
        \includegraphics[width=\linewidth]{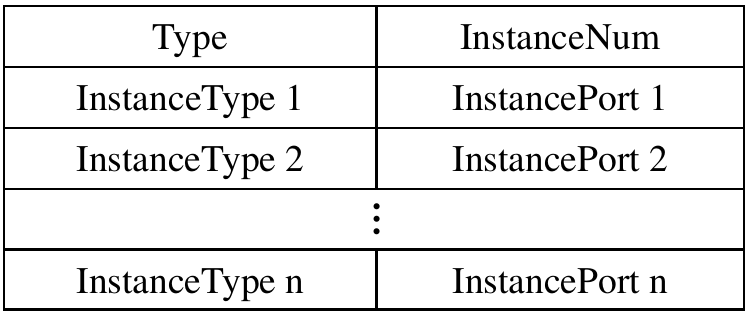}
        \caption{}
        \label{fig7}
    \end{subfigure}
    \hfill % Adds space between the subfigures
    \begin{subfigure}{0.23\textwidth}
        \centering
        \includegraphics[width=\linewidth]{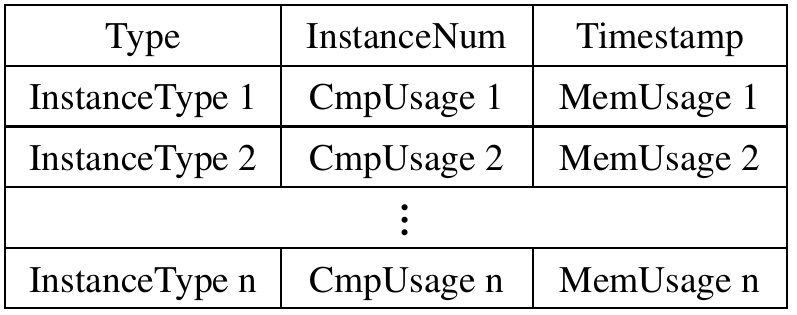}
        \caption{}
        \label{fig8}
    \end{subfigure}
    \vspace{-0.5em}
    \caption{\small The formats of (a) instance update messages and (b) status maintenance messages.}
    \label{fig:combined}
    \vspace{-1.5em}
\end{figure}

 As illustrated in \ref{fig6}, the discovery module first checks whether the target microservice exists locally after acquiring local instance information. If it exists, the discovery module alerts DMSA to initiate monitoring of the instances of the local target microservice. Then, the MA generates and broadcasts instance update messages to other nodes based on the local microservices list, as depicted in Fig. \ref{fig7}, with specific field descriptions provided in Table \ref{tab2}. These instance update messages contain the list of microservice instances on this node and the ports required to invoke the corresponding microservices. After the broadcast, the discovery module enters the listening state and waits for instance update messages from other nodes. Upon receiving such messages, DMSA checks for the presence of target microservices on the corresponding node. If so, it adds these target microservices to the list of available instances and starts monitoring these nodes. Finally, the discovery module determines whether the message is broadcast based on the type field. If so, it responds to these nodes with local instance information; otherwise, no action is taken.

\vspace{-0.5em}

\begin{table}[h]
\setlength{\belowcaptionskip}{-2bp}
\centering
\captionsetup{font=footnotesize} % further reduce font size of caption
\caption{\footnotesize Field descriptions of instance update messages}
\label{tab2}
\renewcommand{\arraystretch}{0.8} % further reduce row height
\setlength{\tabcolsep}{3pt} % further reduce column spacing
\begin{tabular}{ll}
\toprule
Field & Description              \\
\midrule
Type           & Identifier for message type       \\
InstanceNum    & Number of microservice instances  \\
InstancePort $n$ & Agency port for external access   \\
               & of instance $n$                   \\
InstanceType $n$ & Microservice type ID of instance $n$ \\
\bottomrule
\end{tabular}
\vspace{-1em}
\end{table}

\vspace{-1.2em}

\subsection{Implementation of Microservice Monitoring Module}
\par In addition to microservice discovery, the continuous monitoring of instances and prompt sensing of network and node status are indispensable for dynamically adjusting scheduling strategies \cite{zhao2023robust}. However, existing studies usually overlook the dependencies among microservices and only use simple statistical measurements for link status \cite{8990350, vstefanivc2019switch, hannousse2021securing, pahl2016microservices, 8951173}.

\par In this paper, the proposed DMSA monitors three pivotal instance states: CPU and memory usage, along with the available bandwidth of inter-node links, and introduces the corresponding timeout mechanism to guarantee instance availability. Initially, the CPU and memory usage of instances on each node can be directly acquired via the pertinent interfaces furnished by the Docker container engine \cite{8826375,brondolin2020black}. To synchronize this information between MAs on different nodes, status maintenance messages are exchanged. The format of these messages is shown in Fig. \ref{fig8}, with field descriptions in Table \ref{tab3}.

\begin{table}[h]
\setlength{\belowcaptionskip}{-2bp}
\centering
\captionsetup{font=footnotesize} % further reduce font size of caption
\caption{\footnotesize Description of Fields in Status Maintenance Messages}
\label{tab3}
\renewcommand{\arraystretch}{0.8} % further reduce row height
\setlength{\tabcolsep}{2pt} % further reduce column spacing
\begin{tabular}{ll}
\toprule
Field           & Description                          \\
\midrule
Type           & Identifier for message type         \\
InstanceNum    & Number of microservice instances     \\
Timestamp      & Timestamp when the message was sent  \\
InstanceType $n$ & Microservice type ID of instance $n$ \\
CmpUsage $n$     & CPU usage of instance $n$ \\
MemUsage $n$     & Memory usage of instance $n$  \\
\bottomrule
\end{tabular}
\vspace{-0.3em}
\end{table}

\par Since not all microservices are interdependent, monitoring of all instances and nodes is unnecessary. We have considered the dependencies among microservices and proposed a novel synchronization mechanism called PSMS to reduce monitoring overhead, where each MA is both a publisher and a subscriber. As publishers, MAs periodically publish status maintenance messages of their nodes. As subscribers, upon receiving notifications from the discovery module, the monitoring module initiates monitoring processes and subscribes to status maintenance messages from the target node's MA. In addition to synchronization, these status maintenance messages serve as heartbeat packets; if no updates are received from the target node for an extended period, the invocation priority of instances on that node is significantly lowered, thus constituting a timeout mechanism.

\par Furthermore, we develop a measurement mechanism named CAPA to evaluate the link bandwidth and focus on the downlink bandwidth from the target node to the local node due to the large response data volume. To avoid unnecessary overhead, active measurements are primarily performed during initialization or when bandwidth status is outdated. Specifically, the local node sends an active measurement request to the target node, which then returns a fixed-size test data packet $L$. The local node starts timing when the first part of the test data packet arrives and stops timing upon full receipt, with bandwidth calculated as $\frac{L}{t}$\footnote{Without loss of generality, this paper specifies $L$ as $1$ MB. To cope with uplink failures, if the active measurement timeout occurs, the measurement result defaults to $0$ Mb/s.}. The passive measurement is piggybacked by the scheduling module, which records the size $L^{\prime}$ and elapsed time $t^{\prime}$ of each response data transfer. If $L^{\prime}$ exceeds $L$, the status of the node is updated with the passive measurement $\frac{L^{\prime}}{t^{\prime}}$. To reduce measurement errors caused by network fluctuations, we introduce the exponential moving average (EMA) algorithm \cite{cai2021exponential,NAKANO2017187}. It calculates the moving average in an exponentially decaying manner and assigns greater weight to the recent data, thereby smoothing the measurement more efficiently than traditional moving average algorithms. The EMA algorithm prioritizes the latest data and mitigates the effects of older measurements, thus ensuring accurate and up-to-date bandwidth assessments \cite{cai2021exponential,NAKANO2017187}. Let the $i$-th bandwidth measurement be $V_{i}$ and the smoothed bandwidth estimate be $\bar{V}_{i}$, with a smoothing factor of $\theta$, then we have

\vspace{-0.5em}

\begin{equation}\label{eq1}
   \bar{V}_i = \alpha V_i+(1-\alpha)\bar{V}_{i-1}, \ 0<\alpha<1 \ \text{and} \ i>1,
\end{equation}
where $\bar{V}_{1} = V_{1}$ and $\theta = 2$.

\par
\begin{figure}[t]
\vspace{-0.3em}
\centering
\includegraphics[scale=0.4]{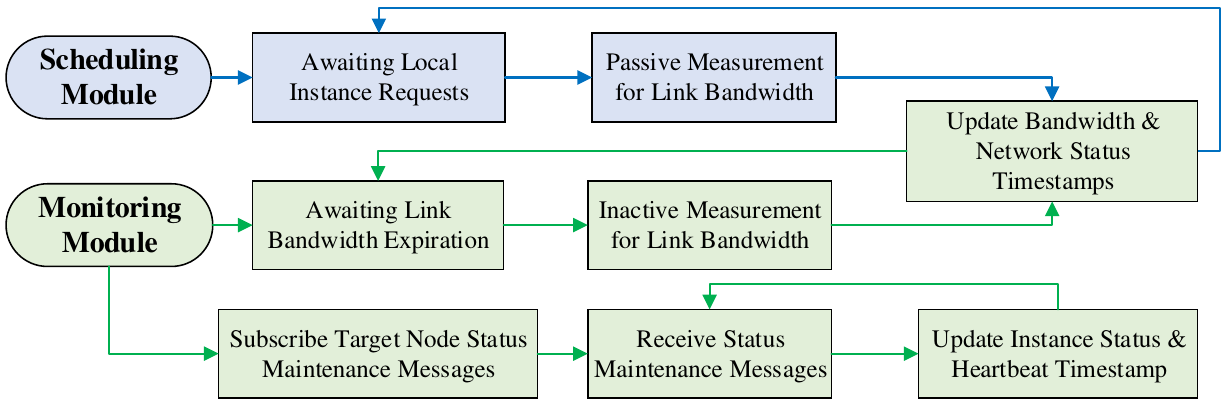}
\vspace{-0.3em}
\caption{\small The flowchart of microservice monitoring process.}
\label{fig9}
\vspace{-0.5em}
\end{figure}

 The monitoring and scheduling modules jointly maintain the target node status table and target instance status table as the primary basis for forwarding requests. The field descriptions are shown in Tables \ref{tab4} and \ref{tab5}. The target node status table includes estimated link bandwidth and heartbeat timestamps, while the target instance status table contains service port and load information for each instance. With these tables, the scheduling module can calculate the priority and weight of each instance to dynamically adjust the scheduling strategies. As illustrated in Fig. \ref{fig9}, upon discovering a new target node, the monitoring module initiates active measurements and subscribes to the node's status maintenance messages. Based on the bandwidth measurements and instance status messages, the new node and its instances are added to the target node status table and target instance status table. After initialization, continuous monitoring ensues. The monitoring module consistently receives status maintenance messages from the target node, updating the node's heartbeat timestamp and the CPU and memory load of its instances. In addition, the monitoring module periodically checks the network status timestamps and reinitiates active measurements when the bandwidth state of the target node has not been updated for an extended period.

\begin{table}[h]
    \setlength{\belowcaptionskip}{-2bp}
    \centering
    \captionsetup{font=footnotesize} % further reduce font size of caption
    \caption{\footnotesize Field Descriptions of Target Node Status Table}
    \label{tab4}
    \renewcommand{\arraystretch}{0.75} % further reduce row height
    \setlength{\tabcolsep}{2pt} % further reduce column spacing
    \begin{tabular}{ll}
        \toprule
        Field          & Description                          \\
        \midrule
        ID             & Target node identifier               \\
        IP             & Node IP address                      \\
        AliveTimestamp & Heartbeat timestamp                  \\
        Bandwidth      & Estimated link bandwidth             \\
        NetTimestamp   & Network status timestamp             \\
        Instances      & List of target instance identifiers  \\
        \bottomrule
    \end{tabular}
    \vspace{-0.5em}
\end{table}

\begin{table}[h]
    \setlength{\belowcaptionskip}{-2bp}
    \centering
    \captionsetup{font=footnotesize} % further reduce font size of caption
    \caption{\footnotesize Field Descriptions of Target Instance Status Table}
    \label{tab5}
    \renewcommand{\arraystretch}{0.8} % further reduce row height
    \setlength{\tabcolsep}{2pt} % further reduce column spacing
    \begin{tabular}{ll}
        \toprule
        Field     & Description                       \\
        \midrule
        ID       & Target instance identifier         \\
        TypeID   & Microservice type ID of instance   \\
        Port     & External proxy port of instance    \\
        CmpUsage & CPU usage of instance              \\
        MemUsage & Memory usage of instance           \\
        NodeID   & Target node ID of instance         \\
        \bottomrule
    \end{tabular}
    \vspace{-1em}
\end{table}

\subsection{Implementation of Microservice Scheduling Module}

\par As the core of DMSA, the scheduling module features two primary functions. Based on the non-intrusive framework, it exploits multi-port listening \cite{bufalino2023analyzing,10542425} and zero-copy forwarding techniques \cite{10179472,10246300} to implement the MA function. Thus, it can take over the external communication of local microservices without modifying the code, correctly distinguish between different invocation types, and ensure efficient forwarding. Secondly, it dynamically computes scheduling strategies in time based on the instance and network status to forward each request to the optimal instance. As depicted in Fig. \ref{fig4}, DMSA configures internal and external IP addresses for communicating with local instances and external nodes, respectively. DMSA listens to multiple ports on these two IP addresses based on the configuration files. The ports for external monitoring correspond one-to-one with local microservices, which are used to forward external requests to these local microservices and track their load. The ports for internal monitoring correspond one-to-one with target microservices, used to forward requests from these local microservices. To use MA, it only needs to specify the destination address of the local microservice to the IP and port corresponding to MA through configuration files or network forwarding rules, leaving the operation logic of the microservice unaffected. As a four-layer proxy, DMSA is beyond caring about the specific content of requests and responses. It directly copies data from the receive buffer to the send buffer, bypassing user space through the zero-copy technique \cite{10179472, 10246300}. This compatibility with any TCP-based application layer protocols significantly improves forwarding efficiency by reducing context switching and CPU load\footnote{The traditional seven-layer proxy working at the application layer involves copying, viewing, and modifying data from the Socket receive buffer to the Socket send buffer repeatedly during request forwarding, and the proxy process necessitates to switch between kernel and user states frequently, which greatly reduces the data forwarding efficiency and increases the CPU load.}. As a result, we customize a dynamic weighted multi-level load balancing (DWMLB) algorithm for DMSA, which comprehensively takes into account the reliability, priority, and response delay to dynamically adjust scheduling strategies based on the instance and node status. In particular, the DWMLB algorithm categorizes instances of each target microservice into three priority levels. The low-priority instances are used only when high-priority ones are unavailable. For instances with the same priority, the DWMLB algorithm assigns different weights based on load and network conditions, favoring more invocations for those instances with lower loads and better network conditions. These instance priorities and weights are dynamically updated to accommodate the complex and varying edge environment. The main variables for the scheduling module are listed in Table \ref{tab6}. Then, we introduce the DWMLB algorithm in detail from three parts: priority division, weight allocation, and weighted scheduling.

\vspace{-0.5em}

\begin{table}[h]
    \setlength{\belowcaptionskip}{6bp}
    \centering
    \caption{\small Main Variables in Microservice Scheduling}
    \label{tab6}
    \scriptsize
    \renewcommand{\arraystretch}{0.85} % reduce row height
    \setlength{\tabcolsep}{4pt} % reduce column spacing
    \begin{tabular}{ll}
    \toprule
    Symbol                         & Definition                         \\
    \midrule
    $M_{i,p}\in \mathcal{M}_i$     & Instance of microservice $M_i$ deployed on node $N_p$        \\
    $C_{i,p}$/$B_{i,p}$            & CPU/memory usage of instance $M_{i,p}$  \\
    $\bar{V}_p$                    & Estimated available link bandwidth to node $N_p$    \\
    $V_p^{max}$                    & Theoretical maximum link bandwidth to node $N_p$          \\
    $T^s_{p}$/$T^c$/$T^{lmt}$      & Heartbeat timestamp of node $N_p$                \\
    $T^c$                          & Current time                        \\
    $T^{lmt}$                      & Timeout threshold for status maintenance messages                   \\
    $C^{lmt}$/$B^{lmt}$/$V^{lmt}$  & CPU max load/memory max load/min bandwidth threshold    \\
    $\alpha$/$\beta$/$\gamma $     & Weights of CPU/memory/bandwidth states in scoring           \\
    $W_{i,q}$ /$W$                 & Scheduling weight/total scheduling weight of instance $M_{i,q}$ \\
    \bottomrule
    \end{tabular}
    \vspace{-0.5em}
\end{table}

\subsubsection{Priority Division} We categorize instances of each target microservice into three priority sequences, i.e., low, medium, and high-priority sequences, and prioritize the high-priority sequences for scheduling. For the set of instances $\mathcal{M}_{i}$ of microservice $M_{i}$, the priority division process is detailed in Algorithm \ref{alg1}, which first checks each instance's heartbeat timestamp for timeouts and places all timeout instances into the low-priority sequence $Q_{i}^l$. When the instances are overloaded, the failure probability increases considerably and the response speed decreases dramatically \cite{9653677}. Consequently, Algorithm \ref{alg1} carefully evaluates the CPU and memory load of the instances, as well as the estimated link bandwidth. Instances exceeding thresholds are placed in the medium-priority sequence $Q_{i}^m$. Instances with normal status are placed in the high-priority sequence $Q_i^h$. Finally, Algorithm \ref{alg1} checks $Q_i^h$, $Q_i^m$, and $Q_i^l$ until the first non-empty sequence is encountered and returned.

\setlength{\textfloatsep}{1pt} % 设置算法与正文之间的间距

\begin{algorithm}
    \caption{\small Target Instance Priority Division Algorithm}%算法名字
    \setstretch{0.3}
    \label{alg1}
    \KwIn{Instance set $\mathcal{M}_{i}$ of target microservice $M_{i}$; Timeout Threshold $T^{lmt}$; Resource thresholds $C^ {lmt}$, $B^{lmt}$, $V^{lmt}$.}

    \KwOut{Highest priority non-empty instance sequence $Q_i$ for target microservice $M_i$.}

    Getting the status of instances and nodes;\\

    Initialize instance sequences $Q_i^h$, $Q_i^m$, $Q_i^l$;\\

    \ForEach{Instance $M_{i,p} \in \mathcal{M}_i$}{

    \If{$T^c-T^s_{p} > T^{lmt}$}{
        Add $M_{i,p}$ to $Q_i^l$;\\
    }

    \ElseIf{$C_{i,p} > C^{lmt}$ or $B_{i,p} > B^{lmt}$ or $\bar{V}_p < V^{lmt}$}{
        Add $M_{i,p}$ to $Q_i^m$;\\
    }

    \Else{
        Add $M_{i,p}$ to $Q_i^h$;\\
    }

    }
    \If{$Q_i^h \neq \emptyset$}{
        \Return $Q_i^h$;\\
    }

    \ElseIf{$Q_i^m \neq \emptyset$}{
        \Return $Q_i^m$;\\
    }

    \Else{
        \Return $Q_i^l$;\\
    }
\end{algorithm}

\setlength{\textfloatsep}{1pt} % 设置算法与正文之间的间距

\subsubsection{Weight Allocation} The weight allocation process for the instance sequence $Q_{i}$ is meticulously outlined in Algorithm \ref{alg2}. Before weight allocation, each instance undergoes a scoring process, which factors in its CPU, memory, and bandwidth idle rates. The CPU and memory idle rates are calculated from the CPU load $C_{i,p}$ and memory load $B_{i,p}$ respectively, while the bandwidth idle rate is based on the estimated link bandwidth $\bar{V}_{p}$. The variables $\alpha$, $\beta$, and $\gamma$ are adjustable to accommodate diverse microservice resource requirements. For compute-intensive services, $\alpha$ and $\beta$ should be increased, while for IO-intensive services \cite{9615028}, a higher $\gamma$ is appropriate. Following scoring, Algorithm \ref{alg2} assigns weights to each instance according to its score for subsequent weighted scheduling. Since weights are typically integers, to reduce errors, it is recommended to set $W$ to a relatively large number, specified as $1000$ in this paper. This dynamic weighting approach significantly enhances the robustness of DMSA. In particular, when an instance is idle, its score and weight increase, thereby taking on more load. Conversely, when overloaded, its score and weight decrease, reducing its invocation probability.

% \vspace{-1em}

\begin{algorithm}
    \caption{\small Target Instance Weight Allocation Algorithm}%算法名字
    \setstretch{0.3}
    \label{alg2}
    \KwIn{Highest priority instance sequence $Q_i$; Instance evaluation parameters $\alpha$, $\beta$, $\gamma$; Total weight $W$}
    \KwOut{Weight set $\mathcal{W}_i$ for instances in $Q_i$.}

    Getting the status of instances and nodes;\\

    Initialize weight $W_{i,p}$ for each instance, score $S_{i,p}$ for each instance, and total score $S$;\\

    \ForEach{$M_{i,p} \in Q_i$}{
    $S_{i,p}$ = $\alpha (1-C_{i,p}) + \beta (1-B_{i,p}) + \gamma \frac{\bar{V}_p}{V_p^{max}}$;\\

    $S+=S_{i,p}$;\\
    }

    \ForEach{$M_{i,p} \in Q_i$}{
        $W_{i,p} =W\frac{S_{i,p}}{S}$;\\

        Update $W_{i,p}$ to $\mathcal{W}_i$;\\
    }
\end{algorithm}

% \vspace{-0.5em}

\subsubsection{Weighted Scheduling} Upon a request targets microservice $M_{i}$, the scheduling module performs the weighted scheduling of $Q_{i}$. The invocation probability of instance $M_{i,p}$ corresponds to its weight $W_{i,p}$. The execution process of weighted scheduling is detailed in Algorithm \ref{alg3}. The system generates a random integer $R$ in the range $[0,W)$. Then it accumulates the weight of each instance in turn until the cumulative sum exceeds $R$, at which point the corresponding instance is returned.

\vspace{-0.3em}

\setlength{\textfloatsep}{1pt} % 设置算法与正文之间的间距

\begin{algorithm}
    \caption{\small Weighted Random Scheduling Algorithm for Target Instances}%算法名字
    \setstretch{0.3}
    \label{alg3}
    \KwIn{Highest priority instance sequence $Q_i$; Instance weight set $\mathcal{W}_i$; Total weight $W$. }
    \KwOut{Selected instance $M_{i,p}$}

    Initialize the current cumulative weight $t=0$;\\

    Randomly generate a number $R$ in the range $[0,W)$;\\

    \ForEach{$M_{i,p} \in Q_i$}{
        t += $W_{i,p}$;\\

        \If{$t>R$}{
            \Return $M_{i,p}$;\\
        }
    }
\end{algorithm}

\vspace{-0.5em}

\par
\begin{figure}[H]
\vspace{-0.2em}
\centering
\includegraphics[scale=0.33]{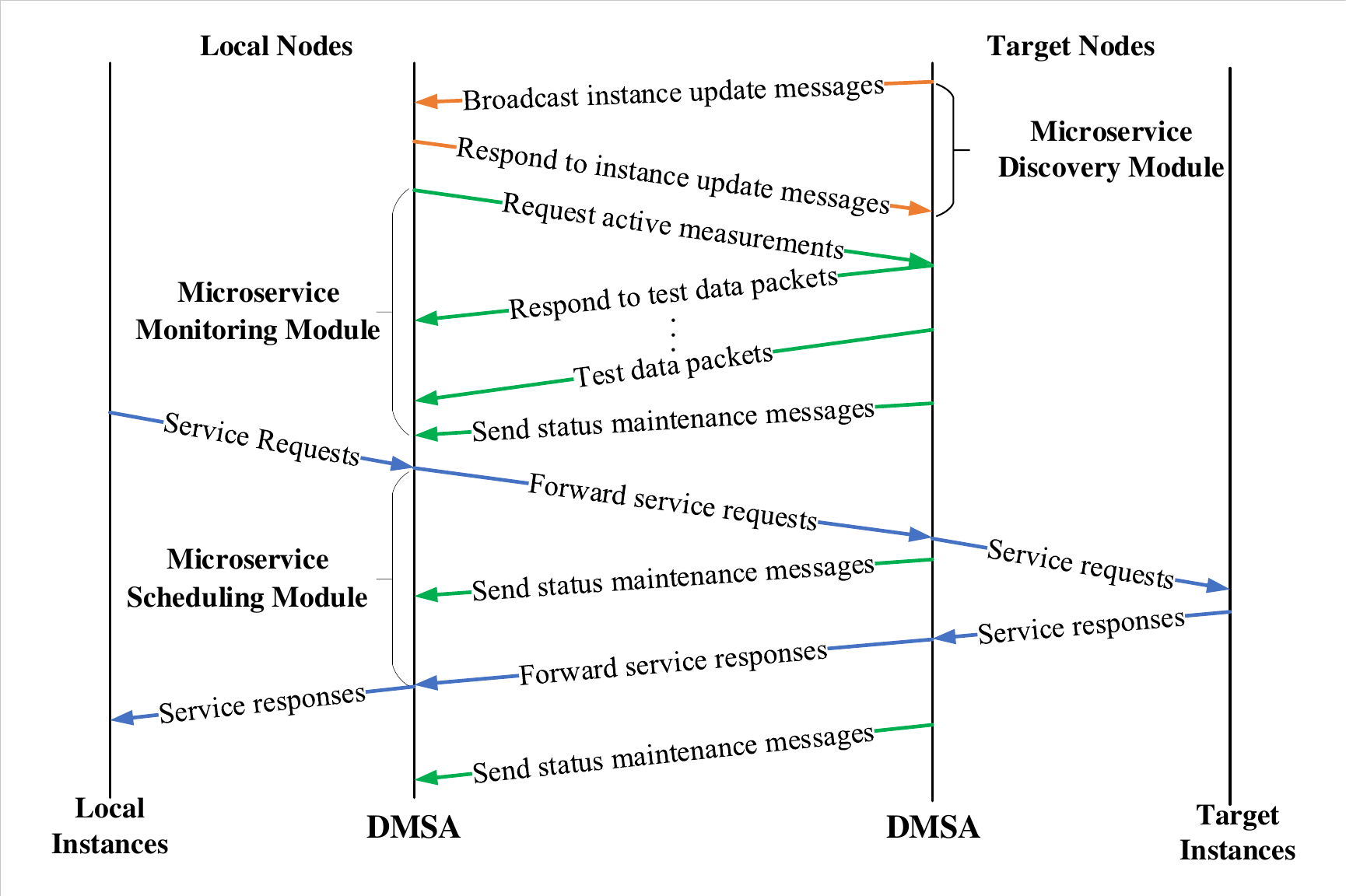}
\vspace{-0.2em}
\caption{\small Data interactions between local nodes and target nodes.}
\label{fig10}
\vspace{-0.5em}
\end{figure}

 The workflow of DMSA is illustrated in Fig. \ref{fig10}. During the initialization, nodes exchange instance update messages to sense each other's instance deployments. The discovery module identifies the presence of target microservices, upon which local microservices depend, based on these instances update messages present on the target node and alert the monitoring module to commence monitoring. This includes proactively measuring link bandwidth status and subscribing to status maintenance messages from the target node. Upon receiving the status of target nodes and target instances, the monitoring module refreshes the target node and instance status tables. Subsequently, the scheduling module dynamically ranks and weights the instances based on these status information. Upon receiving the local instance's request, the scheduling module executes Algorithm \ref{alg3} for the target instance sequence and forwards the request to the corresponding target node port. And the target node's scheduling module routes the request to the target instance. Once the target instance is processed, the response then returns via the same path to the local instance.

\section{Physical Platform Implementation And Performance Evaluation}

\par We establish a physical verification platform for DMSA using multiple \texttt{Raspberry Pi 4B} and \texttt{Orange Pi 5B} devices, interconnected via \texttt{Gigabit Ethernet} switches to create a relatively complex network topology. On this platform, we comprehensively evaluate and analyze the performance of DMSA and compare it with both state-of-the-art and classic schemes to demonstrate its robustness and scalability.

\begin{table}[h]
    \captionsetup{font=small}
    \setlength{\belowcaptionskip}{2bp}
    \centering
    \caption{\small Raspberry Pi 4B and Orange Pi 5B Specifications}
    \label{tab7}
%    \footnotesize
%    \setlength{\tabcolsep}{2.5pt}
%    \begin{tabular}{ll}
    \scriptsize
    \renewcommand{\arraystretch}{0.9} % reduce row height
    \setlength{\tabcolsep}{5pt} % reduce column spacing
    \begin{tabular}{ll}
        \toprule
        Hardware   & Specification                      \\
        \midrule
        Raspberry Pi 4B & \\
        \midrule
        Processor  & Broadcom BCM2711 Quad-core \\
        Memory     & 4 GB DDR4          \\
        USB Ports  & 2x USB 3.0, 2x USB 2.0    \\
        Ethernet   & Gigabit Ethernet                   \\
        \midrule
        Orange Pi 5B & \\
        \midrule
        Processor  & Rockchip RK3588S Octa-core                            \\
        Memory     & 8 GB DDR4                                             \\
        USB Ports  & 1x USB 3.0, 2x USB 2.0,                              \\
                   & 1x USB3.0 Type-C interface                    \\
        Ethernet   & Gigabit Ethernet                                      \\
        \bottomrule
    \end{tabular}
\end{table}

\vspace{-0.5em}

\subsection{Implementation Design for Physical Verification Platform}
\par
\begin{figure}[H]
\vspace{-0.2em}
\centering
\includegraphics[scale=0.16]{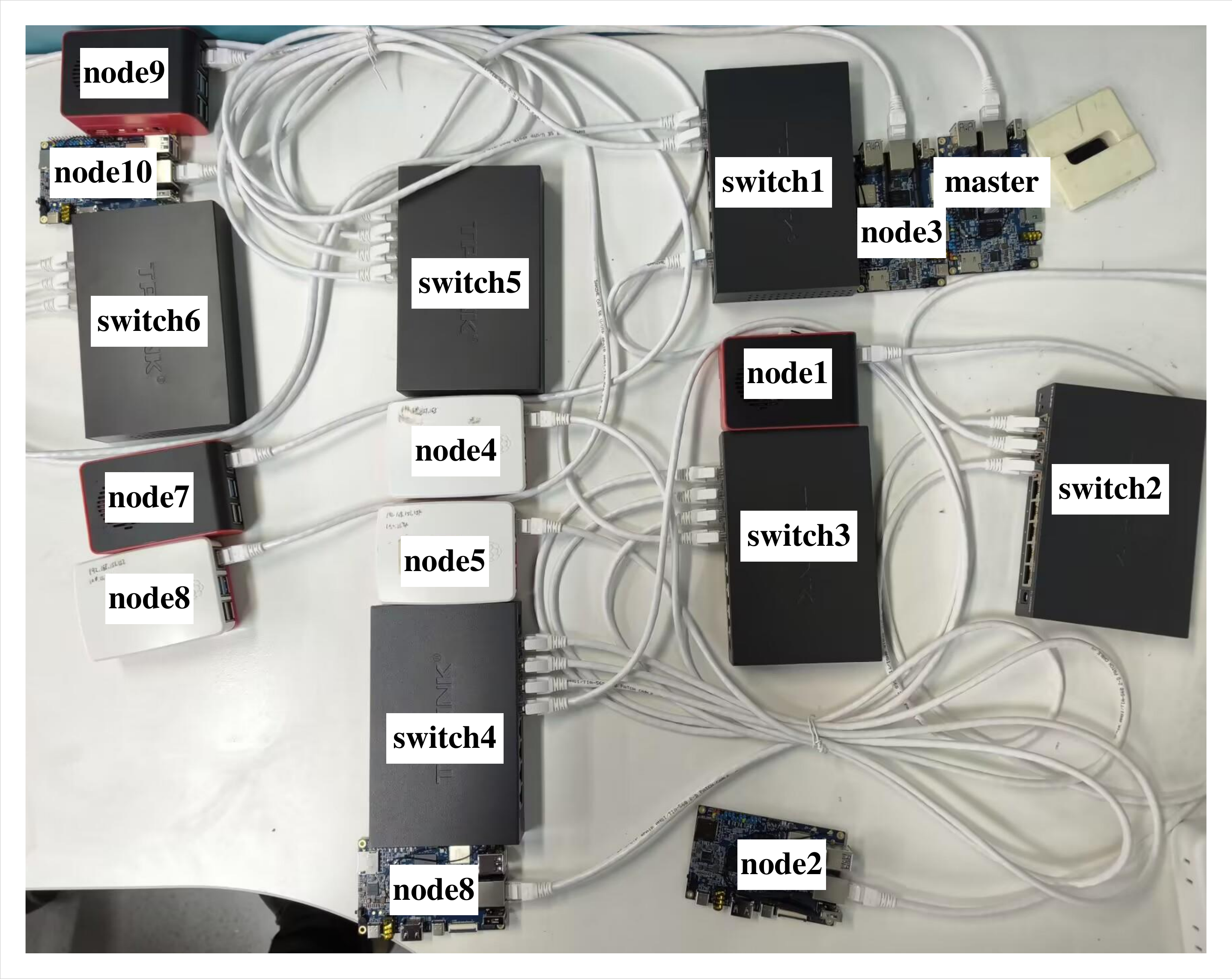}
\vspace{-0.2em}
\caption{\small The physical platform of DMSA.}
\label{fig11}
\vspace{-0.5em}
\end{figure}

\begin{figure}[H]
\vspace{-0.2em}
\centering
\includegraphics[scale=0.3]{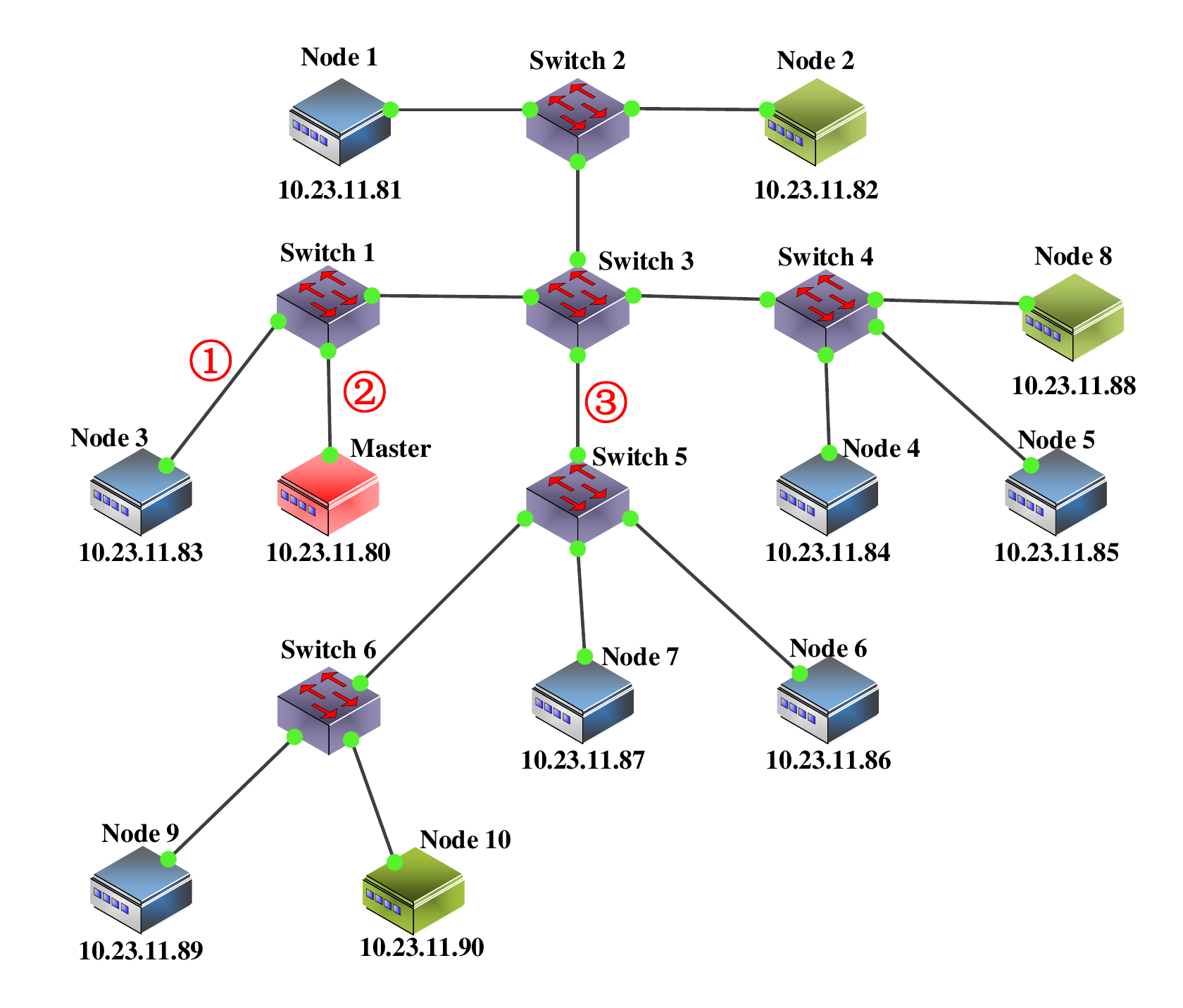}
\vspace{-0.2em}
\caption{\small The logical topology of DMSA.}
\label{fig12}
\vspace{-0.5em}
\end{figure}

 This DMSA platform comprises 6 \texttt{Raspberry Pi 4B}s, 5 \texttt{Orange Pi 5B}s, and 6 \texttt{Gigabit Ethernet} switches, with their parameters detailed in Tables \ref{tab7}. The hardware setup and its logical topology are depicted in Fig. \ref{fig11} and Fig. \ref{fig12}, respectively. The constructed network topology includes 17 nodes: 6 \texttt{Gigabit Ethernet} switches as communication nodes, 3 \texttt{Raspberry Pi 4B}s as user-access nodes (nodes 2, 8, and 10), 1 \texttt{Raspberry Pi 4B} as the Master node, and 7 compute nodes (2 \texttt{Raspberry Pi 4B}s and 5 \texttt{Orange Pi 5B}s). Notably, the Master node acts as a compute node in DMSA and as a control center in centralized MSAs.

\par
\begin{figure}[H]
\vspace{-0.2em}
\centering
\includegraphics[scale=0.45]{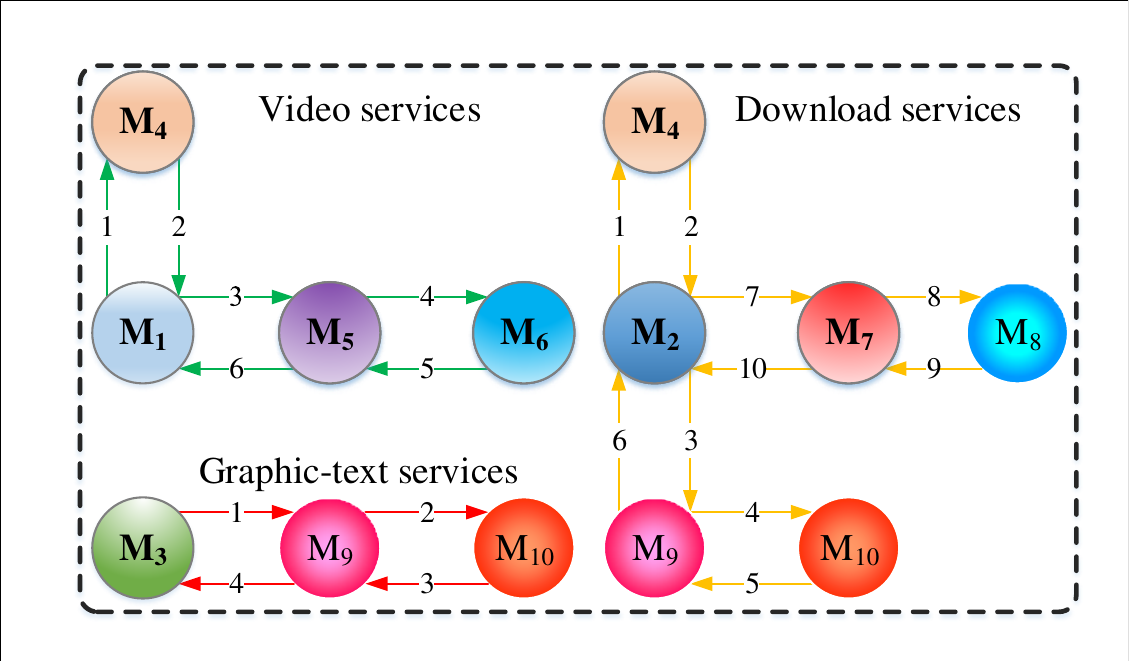}
\vspace{-0.2em}
\caption{\small Service types and microservice invocation relationships.}
\label{fig13}
\vspace{-0.5em}
\end{figure}

 As depicted in Fig. \ref{fig13}, we design three typical services: video services, download services, and graphic-text services. Video services, including short videos, live streaming, and remote monitoring, are latency-sensitive and bandwidth-intensive, making them ideal for edge deployments \cite{feng2022computation,8924682,9419855}. Download services, requiring high bandwidth for large file transfers, are also edge-deployed due to privacy and security concerns \cite{nain2022towards,kong2022edge}. Graphic-text services, including news sites, forums, and shopping sites, represent the most common web services. These services are implemented by 10 microservices, each with specific functions, dependencies, and instances detailed in Table \ref{tab8}.

\par Then, we detail the experimental procedure and related parameters \cite{9615028, 9993766}. Each link in the topology has a bandwidth of $1000$ Mb/s. Each instance can use up to one CPU core and $1$ GB of memory. Video segments range from 1 to 3 MB with a maximum wait time of $10$ seconds. Web pages are 0.5 to 1 MB with the same wait time, while download files are 10 to 20 MB with a 100-second wait time. If the timeout elapses, the service execution is marked as failed. Three user-access nodes generate requests for the three service types at a specific arrival rate with loads evenly distributed. Each test lasts $40$ minutes. To comprehensively evaluate the performance of DMSA in edge networks, three network events are designed as illustrated in Fig. \ref{fig12}. {\Large \ding{192}} Link $1$ between Switch $1$ and Node $3$ disconnects at the $10$th minute and restores at the $15$th to simulate the compute node suddenly goes offline. {\Large \ding{193}} Link $2$ between Switch $1$ and master disconnects at the 20th minute and restores at the $25$th to evaluate the control center suddenly goes down. {\Large \ding{194}} Link $3$ between Switch $3$ and Switch $5$ is limited to $100$ Mb/s at the $30$th minute and restores at the $35$th to portray the network fluctuations. As detailed in Table \ref{tab9}, we also test the performance of DMSA under high, medium, and low load conditions, continuously measuring service response delay and execution success rate.

\begin{table}[h]
\captionsetup{font=small}
\setlength{\belowcaptionskip}{2bp}
\centering
\caption{\small Microservice Functions and Dependencies}
\label{tab8}
\scriptsize
\renewcommand{\arraystretch}{0.9} % reduce row height
\setlength{\tabcolsep}{5pt} % reduce column spacing
\begin{tabular}{lllc}
\toprule
ID & Function & Dependencies & Instance Counts \\
\midrule
$M_1$ & Video Server & $M_4$, $M_5$ & 3 \\
$M_2$ & File Download Server & $M_4$, $M_7$, $M_9$ & 3 \\
$M_3$ & Web Server & $M_9$ & 3 \\
$M_4$ & User Authentication Center & None & 6 \\
$M_5$ & Video Content Distribution & $M_6$ & 3 \\
$M_6$ & Video Content Center & None & 1 \\
$M_7$ & File Cache & $M_8$ & 3 \\
$M_8$ & File Storage Center & None & 1 \\
$M_9$ & Database Middleware & $M_{10}$ & 6 \\
$M_{10}$ & Database & None & 1 \\
\bottomrule
\end{tabular}
\vspace{-0.5em}
\end{table}

\begin{table}[h]
\captionsetup{font=small}
\setlength{\belowcaptionskip}{2bp}
\centering
\caption{\small Arrival Rates of Different Services under Various Loads}
\label{tab9}
%\footnotesize
%\setlength{\tabcolsep}{3pt}
\scriptsize
\renewcommand{\arraystretch}{0.9} % reduce row height
\setlength{\tabcolsep}{5pt} % reduce column spacing
\begin{tabular}{lccc}
\toprule
Load Condition & Video Service & Download Service & Graphic-text Service \\
\midrule
High Load & 5 & 1 & 10 \\
Medium Load & 3 & 0.6 & 6 \\
Low Load & 1.5 & 0.3 & 3 \\
\bottomrule
\end{tabular}
\vspace{-0.5em}
\end{table}

\subsection{Baseline Schemes for Comparison}
\par To thoroughly demonstrate the superiority of DMSA, we compare it with the following state-of-the-art and classical scheduling schemes:

\begin{itemize}
  \item Nautilus \cite{9615028, 9460542, 10592806, 9820678}: As a state-of-the-art centralized microservice scheduling scheme, Nautilus prioritizes network load, CPU, and memory allocation of edge nodes. The central scheduler dynamically adjusts scheduling strategies based on continuous monitoring of each node's load.

  \item Least Connects (LC) algorithm \cite{yang2024reducing}: This baseline scheme tacks real-time connections to each instance, selecting the one with the fewest connections. However, it cannot sense changes in the status of nodes and instances timely since it lacks inter-node status exchange.

  \item Round Robin (RR) scheduling algorithm \cite{nasser2016analisis}: An extensively used static method, RR cycles through target instances, invoking each in turn.
\end{itemize}

\vspace{-0.5em}

\subsection{Empirical Results Measured over DMSA Platform}

\begin{figure}[t]
    \centering
    \begin{minipage}[b]{0.5\textwidth}
        \centering
        \begin{subfigure}{0.8\columnwidth}
            \centering
            \includegraphics[width=\linewidth]{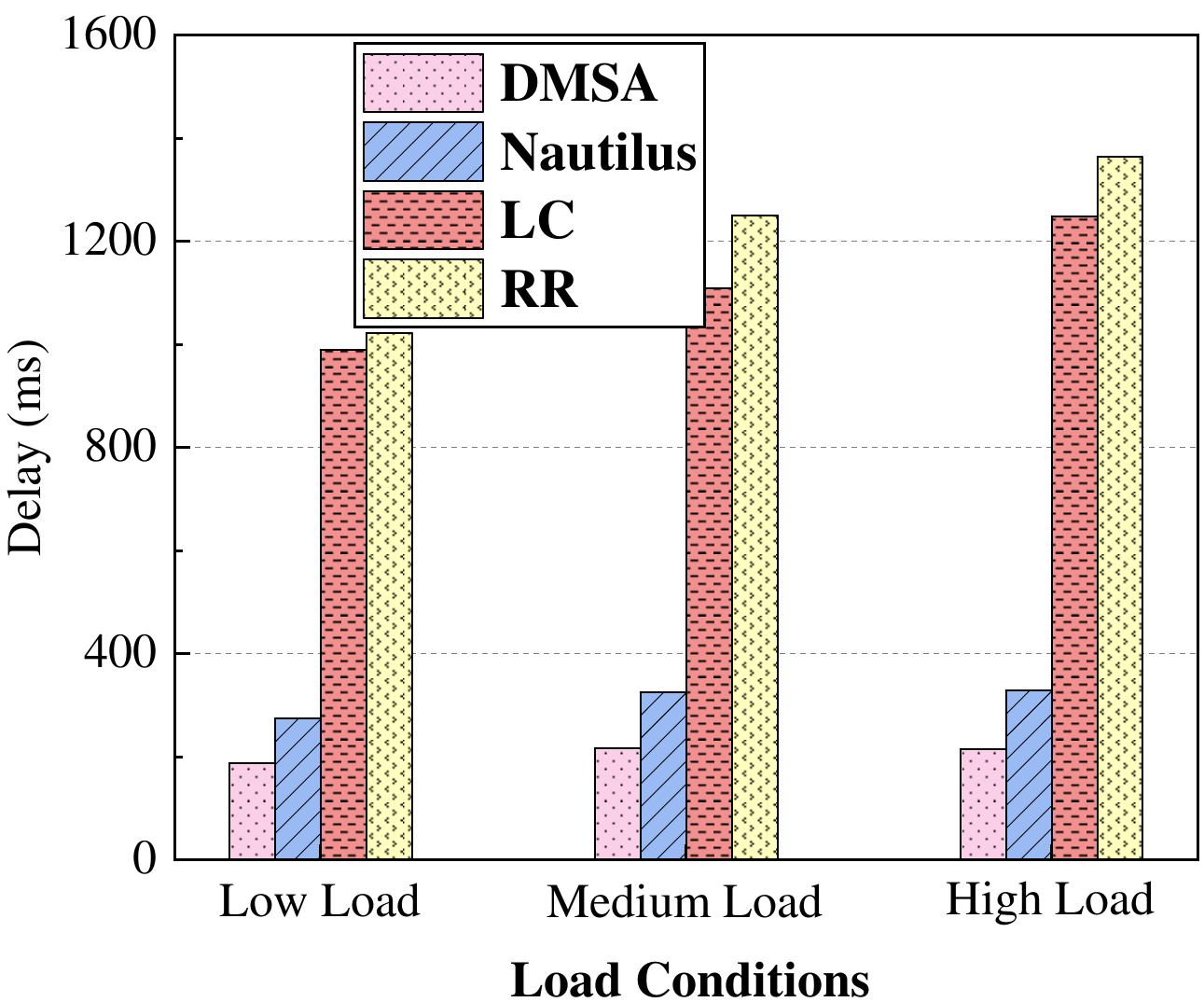}
            \caption{Graphic-text Services}
        \end{subfigure}
        \hfill
        \begin{subfigure}{0.8\columnwidth}
            \centering
            \includegraphics[width=\linewidth]{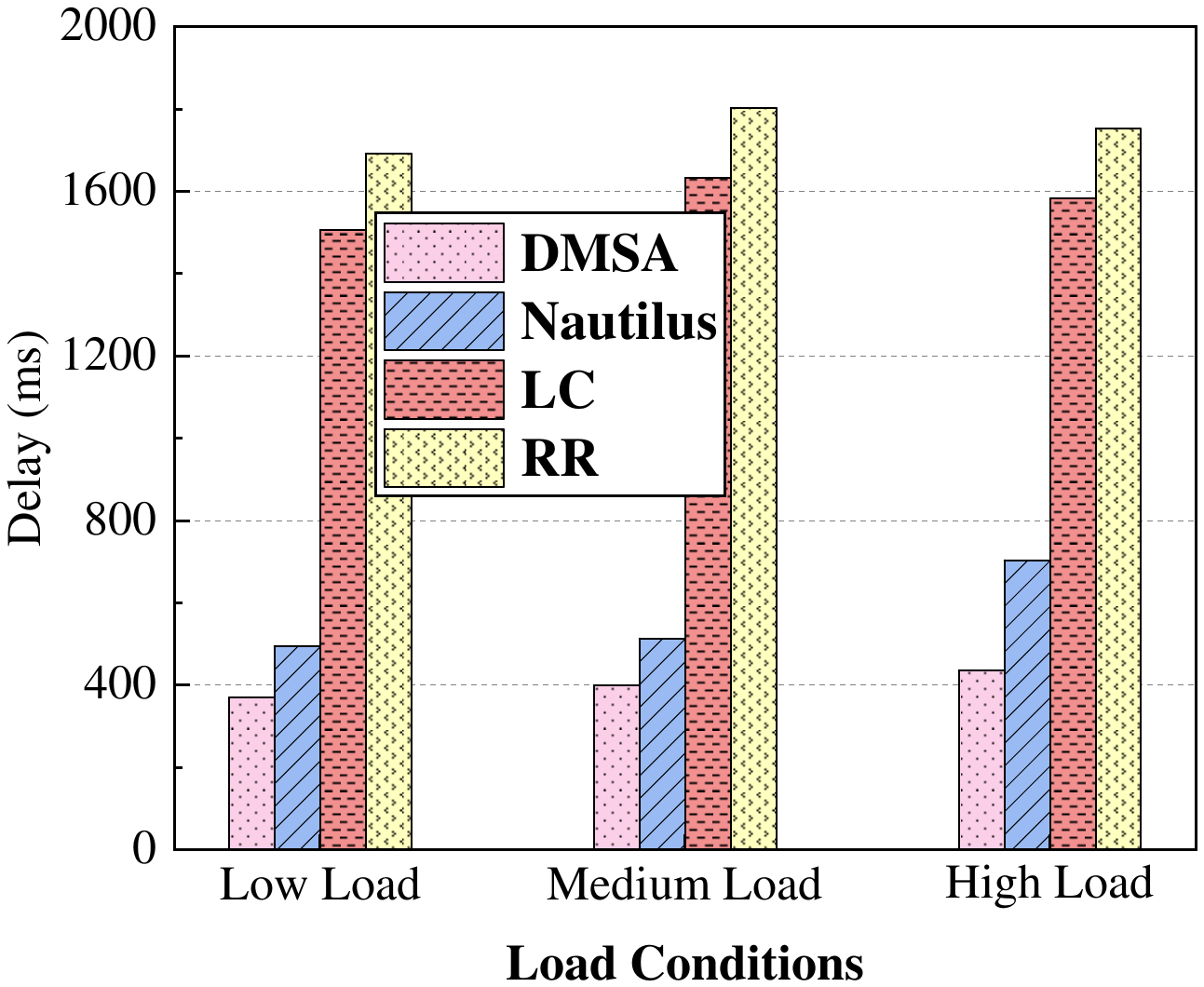}
            \caption{Video Services}
        \end{subfigure}
        \vspace{-0.3em}
        \caption{\small Average response delay for graphic-text and video services under different load conditions.}
        \vspace{0.3em}
        \label{fig14}
    \end{minipage}
    \vspace{-0.3em}
\end{figure}

\subsubsection{Performance Analysis of DMSA for Latency-Sensitive Services}

\par For latency-sensitive services like video and graphic text, we focus on response delay performance. As shown in Fig. \ref{fig14}, average response delays under various load conditions reveal that Nautilus slightly outperforms LC and RR. However, our proposed DMSA excels in handling network fluctuations, significantly outperforming the baseline schemes. In particular, DMSA reduces average response delay by approximately 60\% to 75\% compared to Nautilus, LC, and RR. These real-world measurements underscore DMSA's superior performance in microservice scheduling and also reveal the flexibility and developer-friendly nature of further evolving microservices scheduling schemes on our DMSA platform.

\subsubsection{Performance Analysis of DMSA under Different Network Events}

\par
\begin{figure*}[t]
    \centering
    \begin{minipage}[b]{0.9\textwidth}  % 调整整体宽度
        \centering
        \begin{subfigure}{0.325\textwidth}  % 每个子图的宽度调整为1/3
            \centering
            \includegraphics[width=\linewidth]{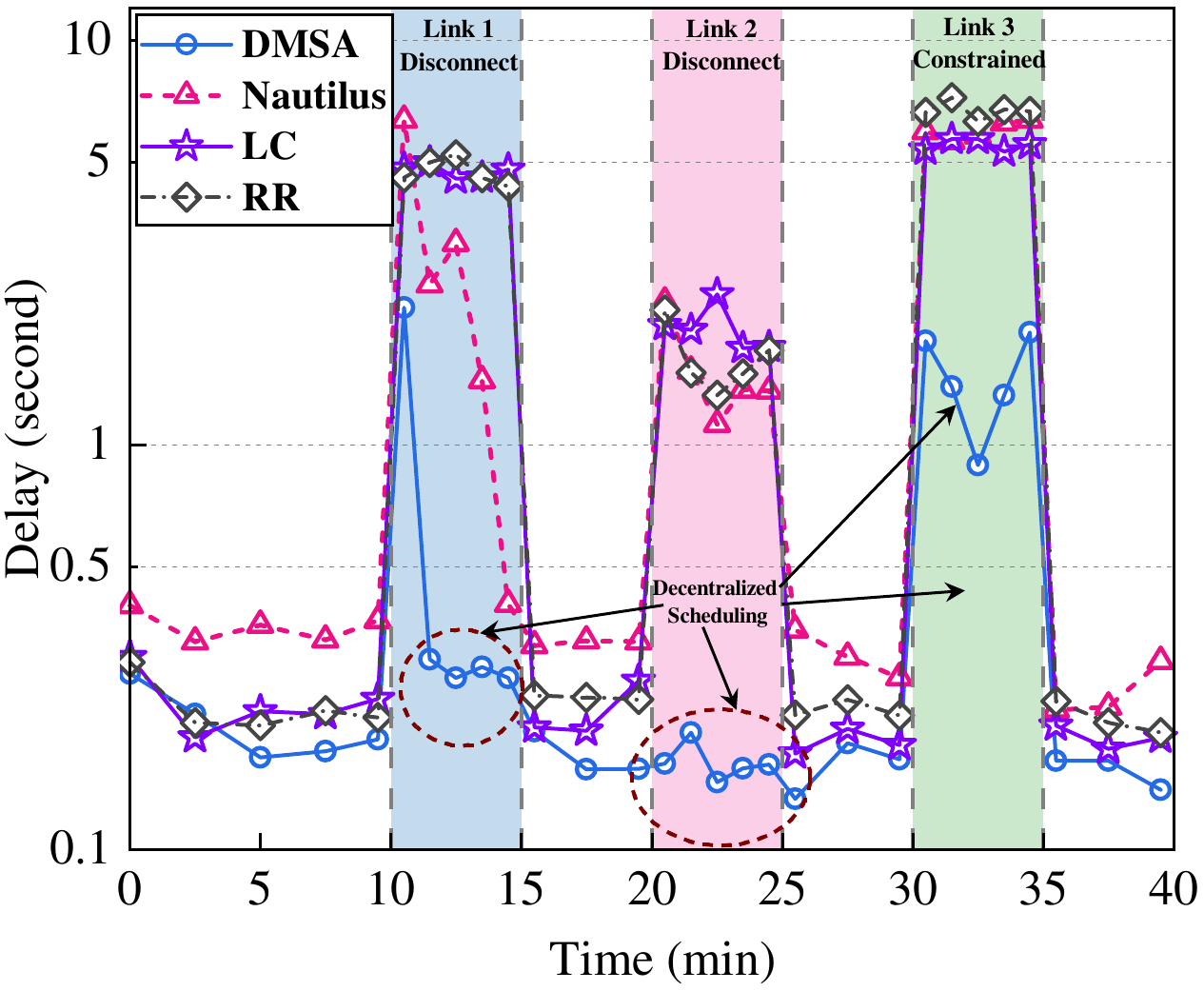}
            \caption{Low load: response delay}
        \end{subfigure}
        \hfill
        \begin{subfigure}{0.325\textwidth}
            \centering
            \includegraphics[width=\linewidth]{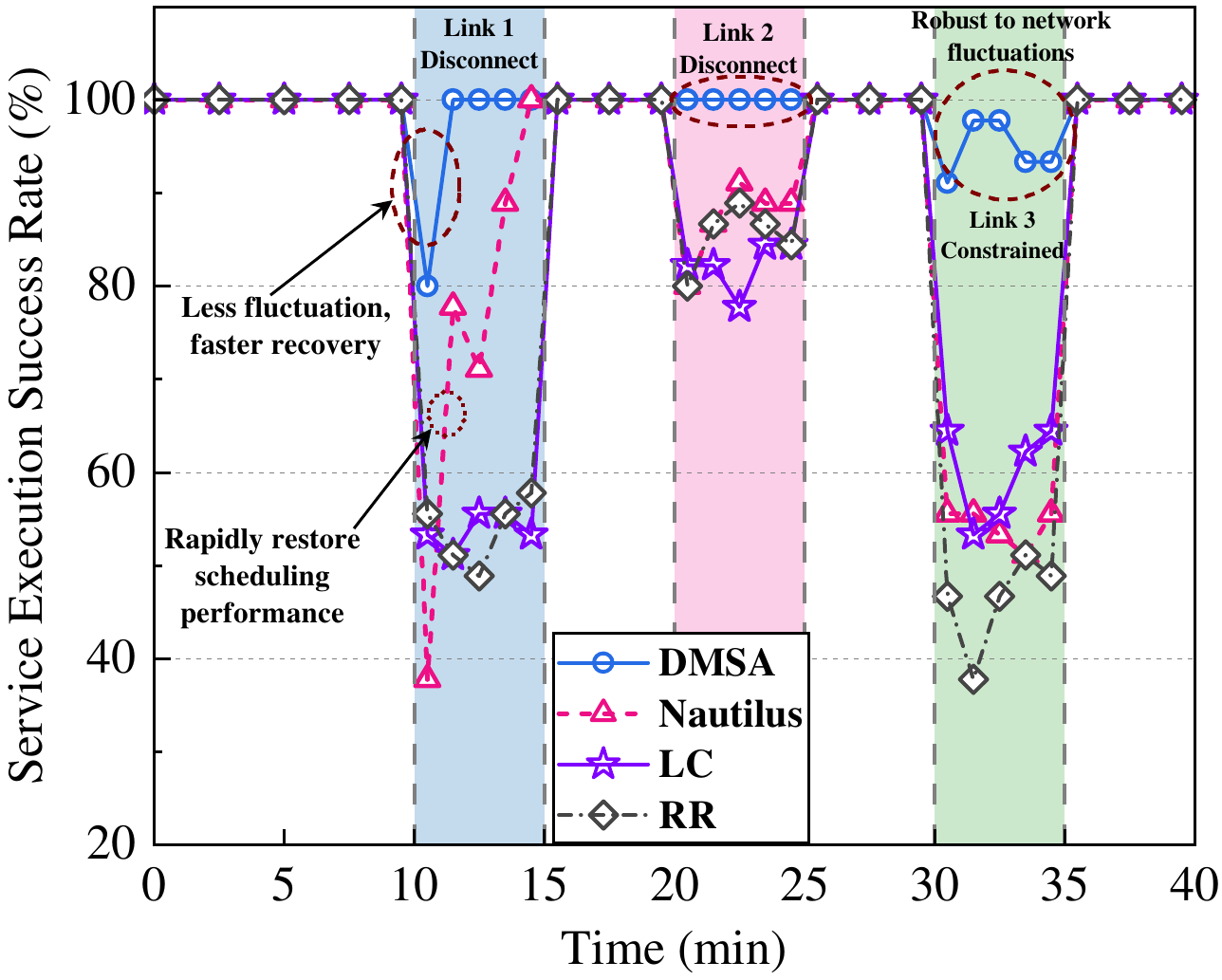}
            \caption{\small Low load: success rate}
        \end{subfigure}
        \hfill
        \begin{subfigure}{0.325\textwidth}
            \centering
            \includegraphics[width=\linewidth]{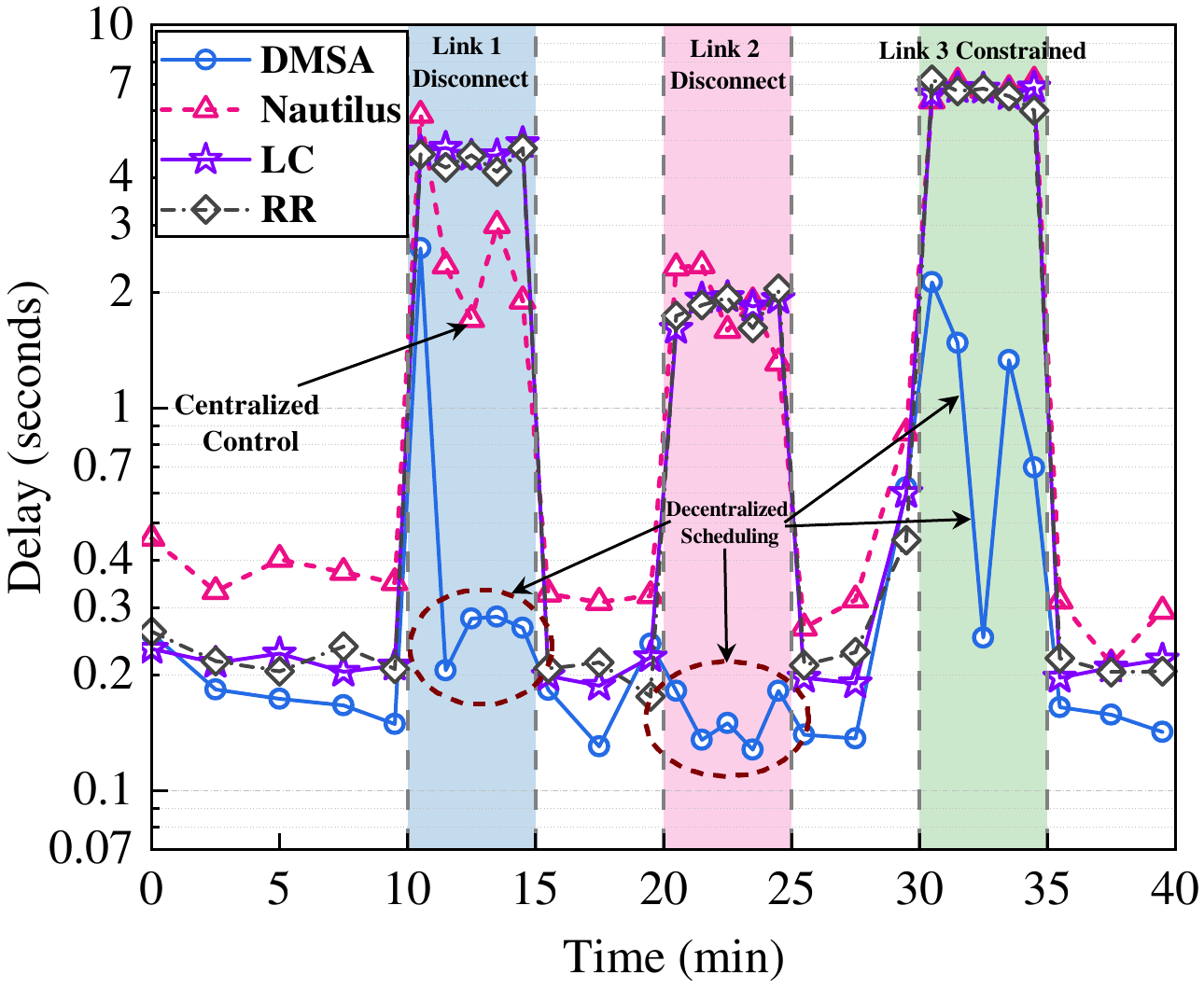}
            \caption{\small Medium load: response delay}
        \end{subfigure}

        \vspace{0.5em}  % 控制行间距

        \begin{subfigure}{0.325\textwidth}
            \centering
            \includegraphics[width=\linewidth]{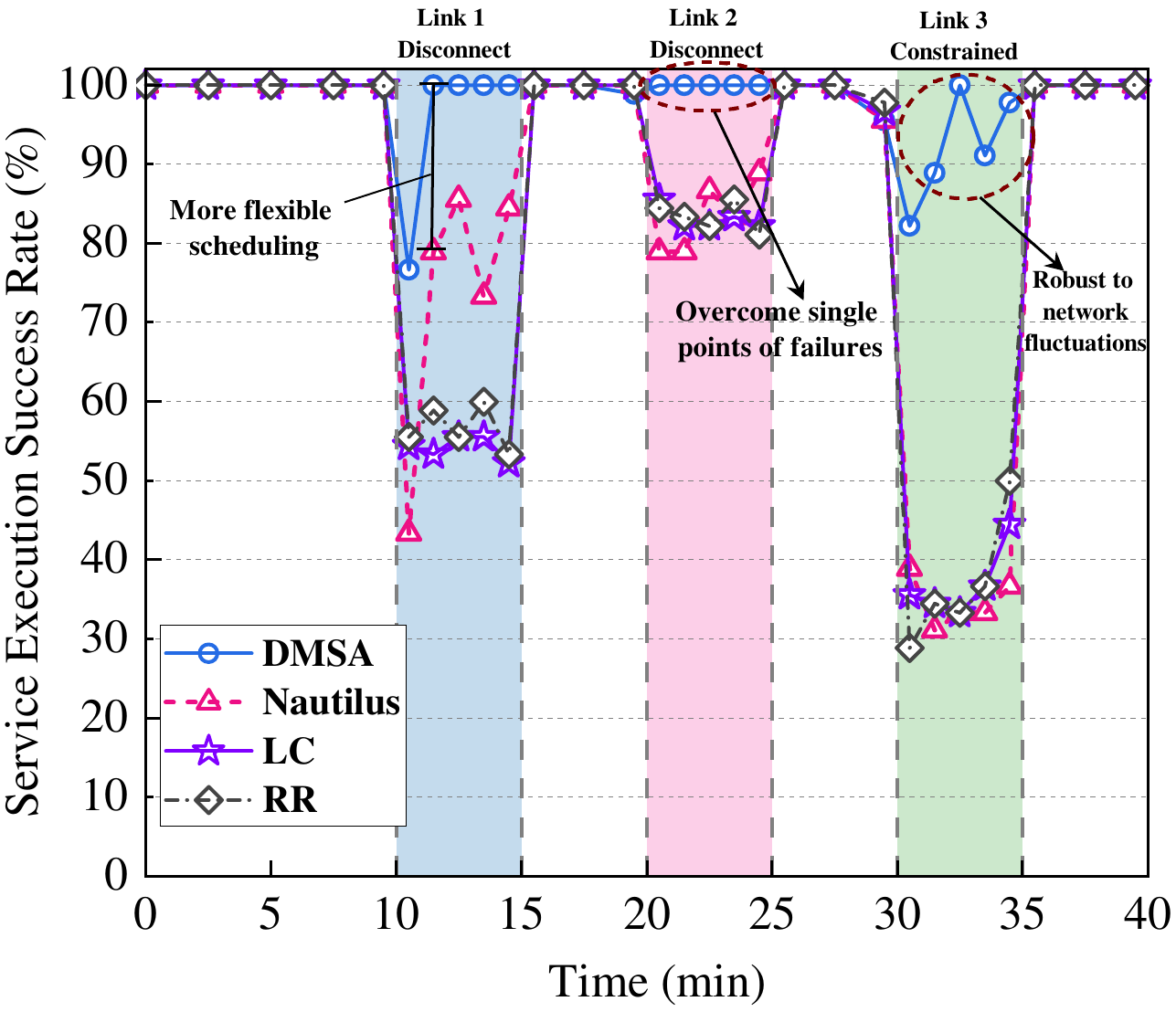}
            \caption{\small Medium load: success rate}
        \end{subfigure}
        \hfill
        \begin{subfigure}{0.325\textwidth}
            \centering
            \includegraphics[width=\linewidth]{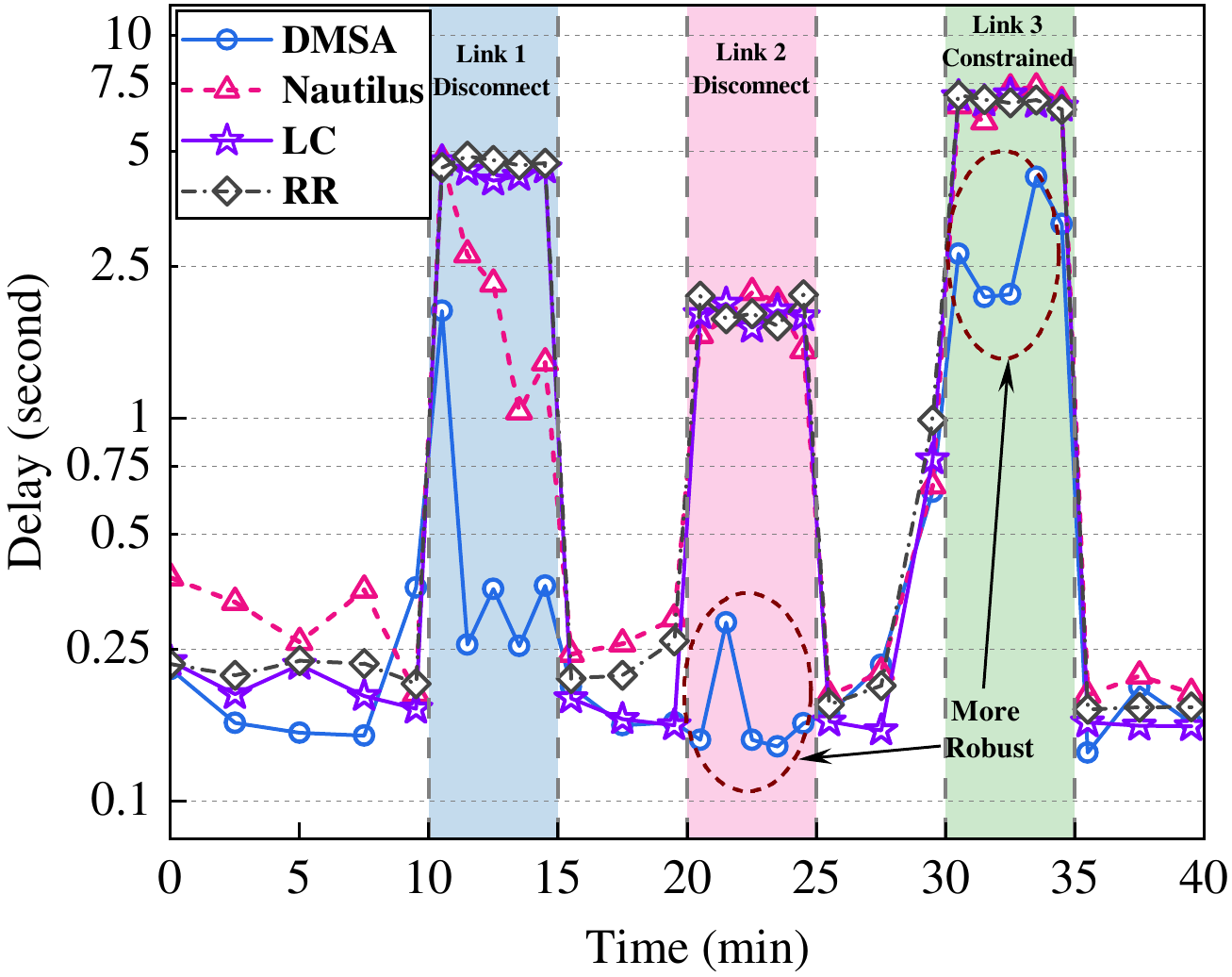}
            \caption{\small High load: response delay}
        \end{subfigure}
        \hfill
        \begin{subfigure}{0.325\textwidth}
            \centering
            \includegraphics[width=\linewidth]{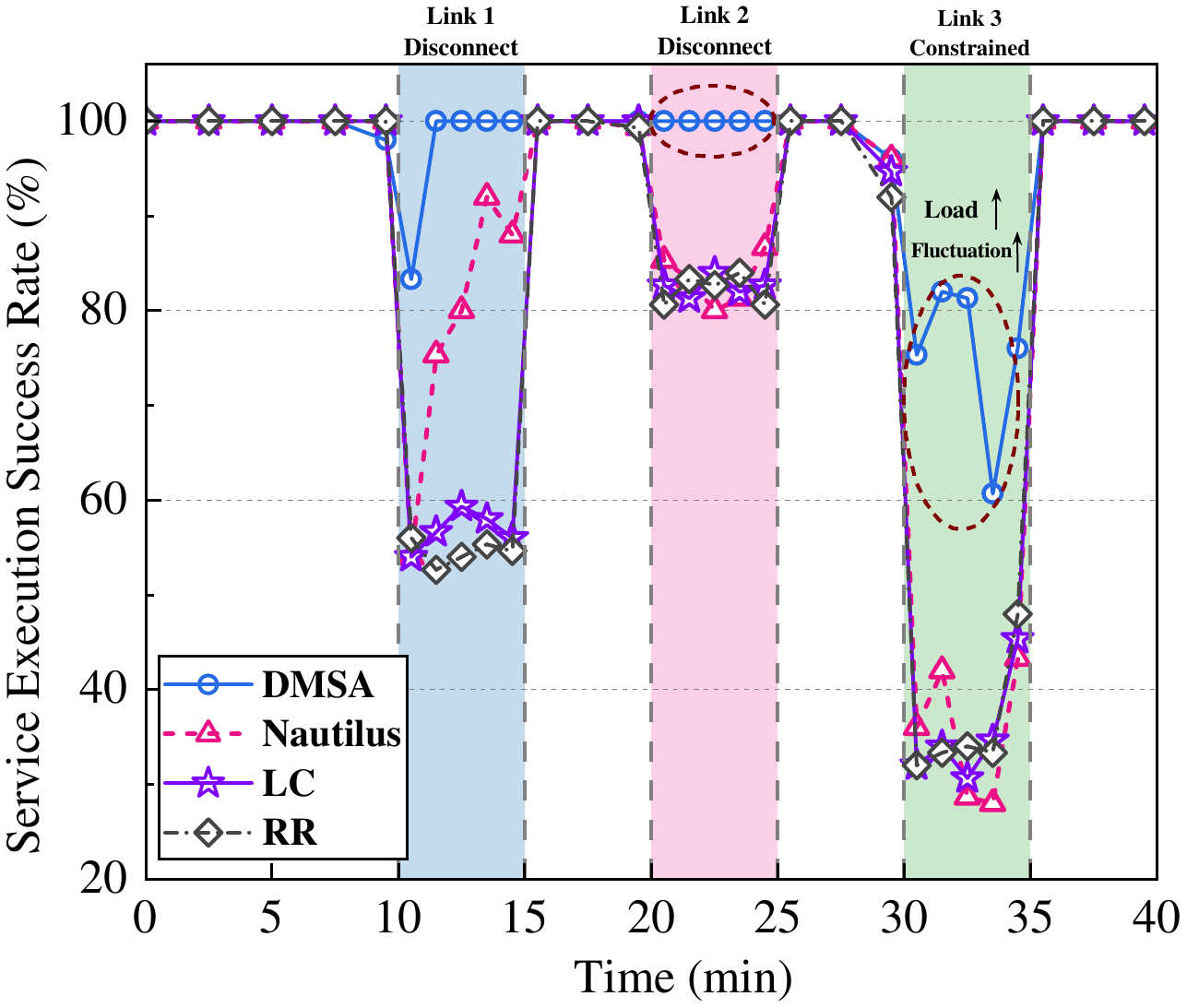}
            \caption{\small High load: success rate}
        \end{subfigure}
        \vspace{-0.3em}
        \caption{\small The dynamic variation of response delay and service execution success rate for video services under different load conditions.}
        \label{fig15}
    \end{minipage}
    \vspace{-1em}
\end{figure*}

 Taking video services as an example, Fig. \ref{fig15} illustrates the dynamic variations in response delay and service execution success rate under different load conditions. Fig. \ref{fig15} (a), (c), and (e) demonstrate that DMSA significantly outperforms other baseline schemes in response delay across diverse loads. A significant network disruption occurs at {\large \ding{192}} $10 \sim 15$ minutes, where Link 1 between Switch 1 and Node 3 disconnects. Fig. \ref{fig15} reveals that the performance of LC and RR algorithms sharply deteriorates and fails to recover promptly due to their lack of awareness about network and node status. This results in severe congestion as user requests to be continuously forwarded to Node 3 despite the link failure. In contrast, Nautilus quickly recovers the network's performance from this link failure. This is mainly because Nautilus effectively perceives network load and edge node's status, dynamically adjusting scheduling strategies based on continuous monitoring after detecting Link 1's disconnection, stopping forwarding user requests to Node 3. Remarkably, DMSA exhibits even faster recovery capability. Attributed to DMSA's efficient discovery and monitoring modules, MAs interact through configuration files and instance update messages to adeptly pinpoint topological changes, as depicted in Fig. \ref{fig5} and \ref{fig6}. Concurrently, the PSMS mechanism significantly reduces monitoring overhead. Combining active and passive measurements with the EMA algorithm substantially mitigates network fluctuation-induced measurement errors, guaranteeing precise and prompt monitoring, as shown in Fig. \ref{fig9}.

\begin{figure*}[t]
    \centering
    \begin{subfigure}{0.6\columnwidth}
        \centering
        \includegraphics[width=\linewidth]{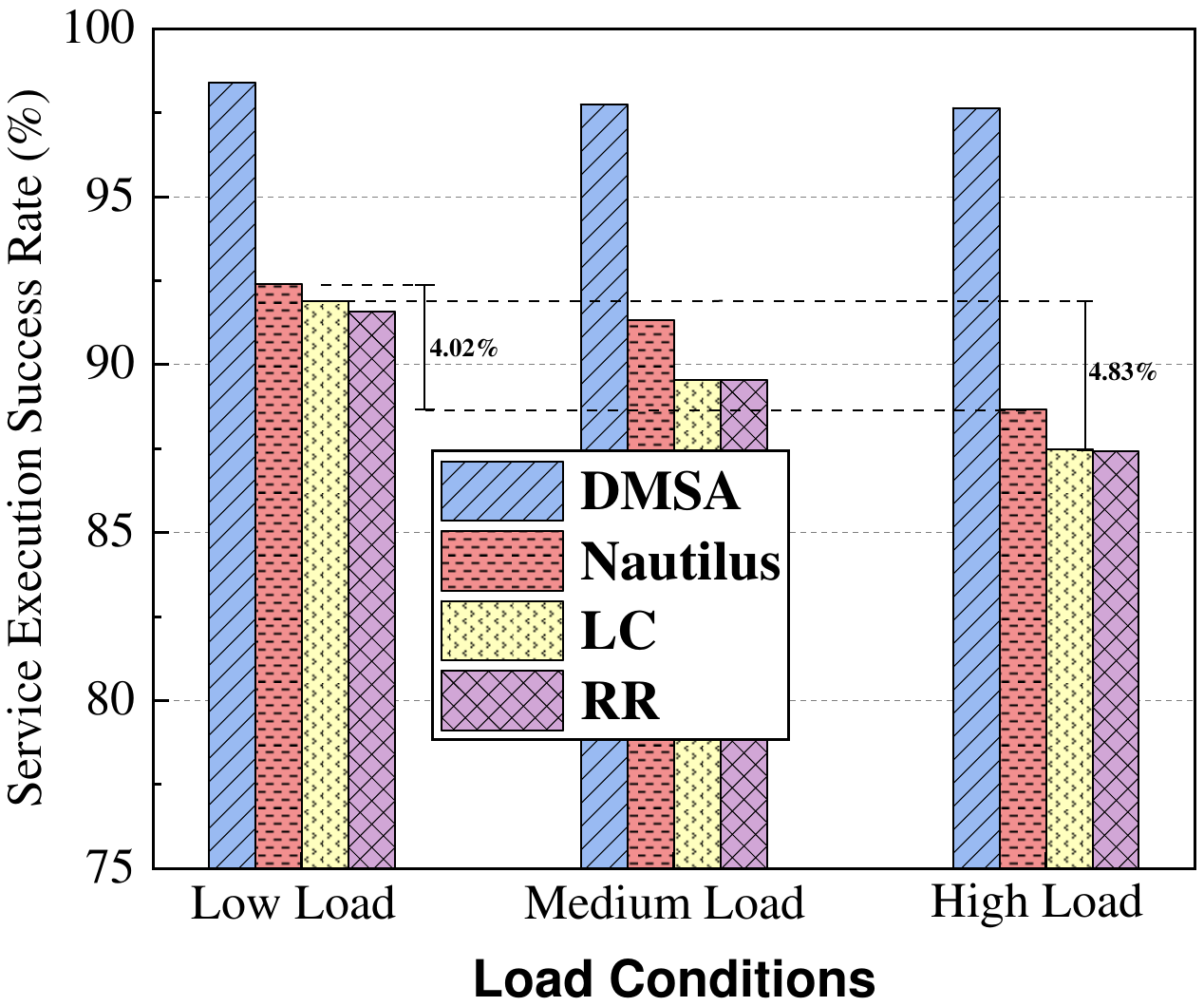}
        \caption{\small Graphic-Text Services}
    \end{subfigure}
    \hspace{1em}
    \begin{subfigure}{0.6\columnwidth}
        \centering
        \includegraphics[width=\linewidth]{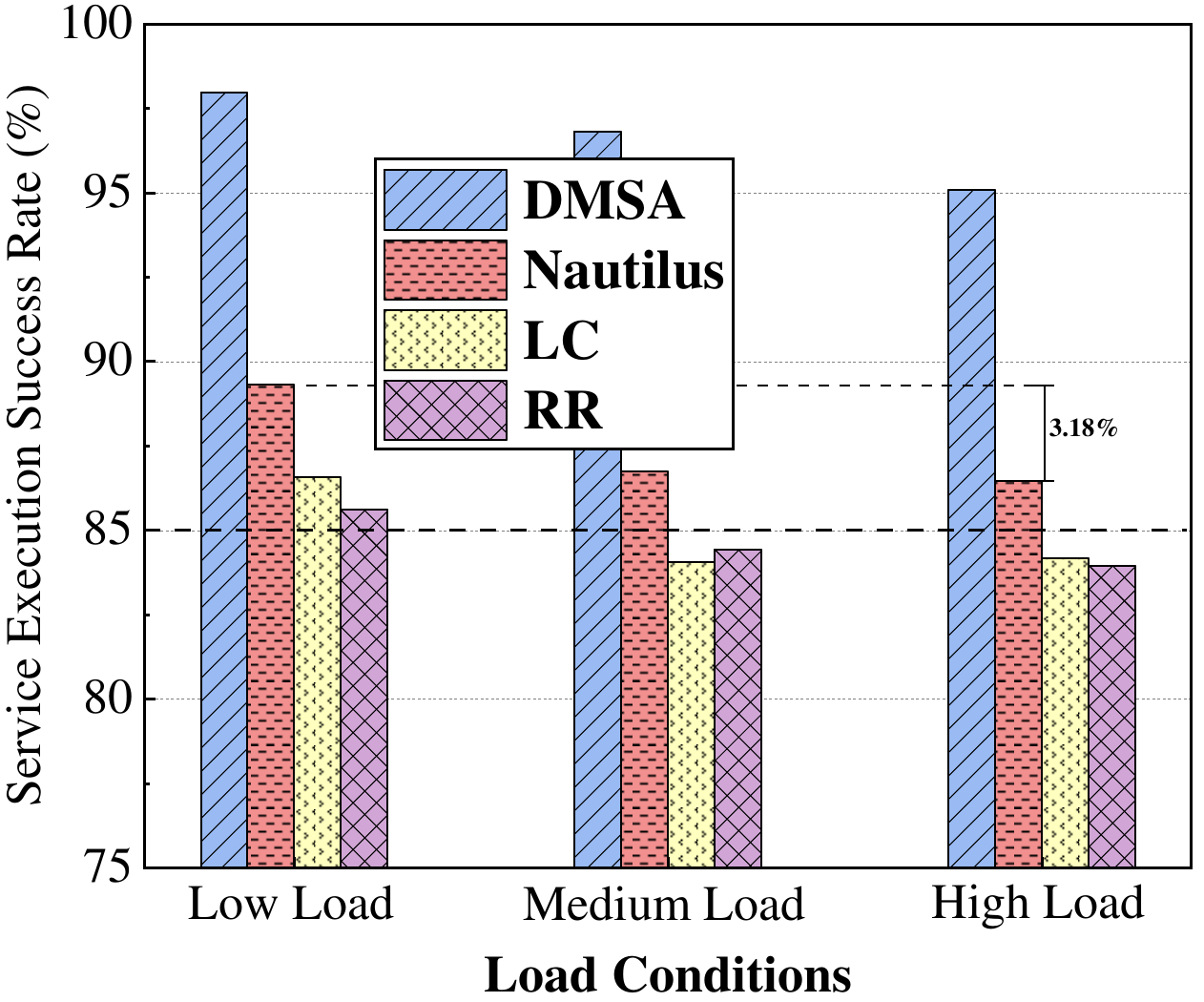}
        \caption{\small Video Services}
    \end{subfigure}
    \hspace{1em}
    \begin{subfigure}{0.6\columnwidth}
        \centering
        \includegraphics[width=\linewidth]{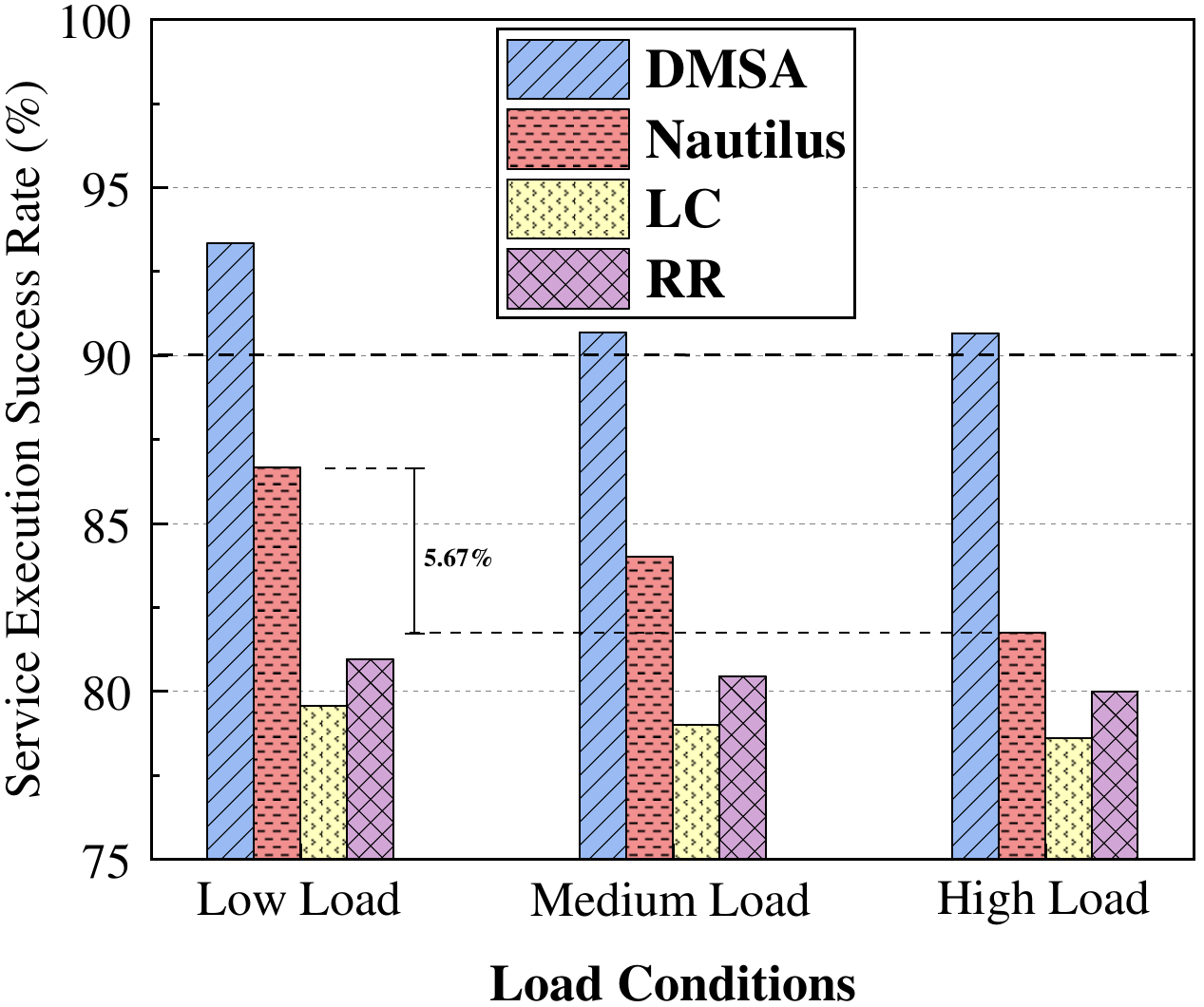}
        \caption{\small File Download Services}
    \end{subfigure}
    \caption{Service execution success rates for services under different load conditions.}
    \label{fig16}
    \vspace{-1em}
\end{figure*}

\begin{figure}[h]
    \centering
    \begin{minipage}[b]{0.5\textwidth}
        \centering
        \begin{subfigure}{0.9\columnwidth}
            \centering
            \includegraphics[width=\linewidth]{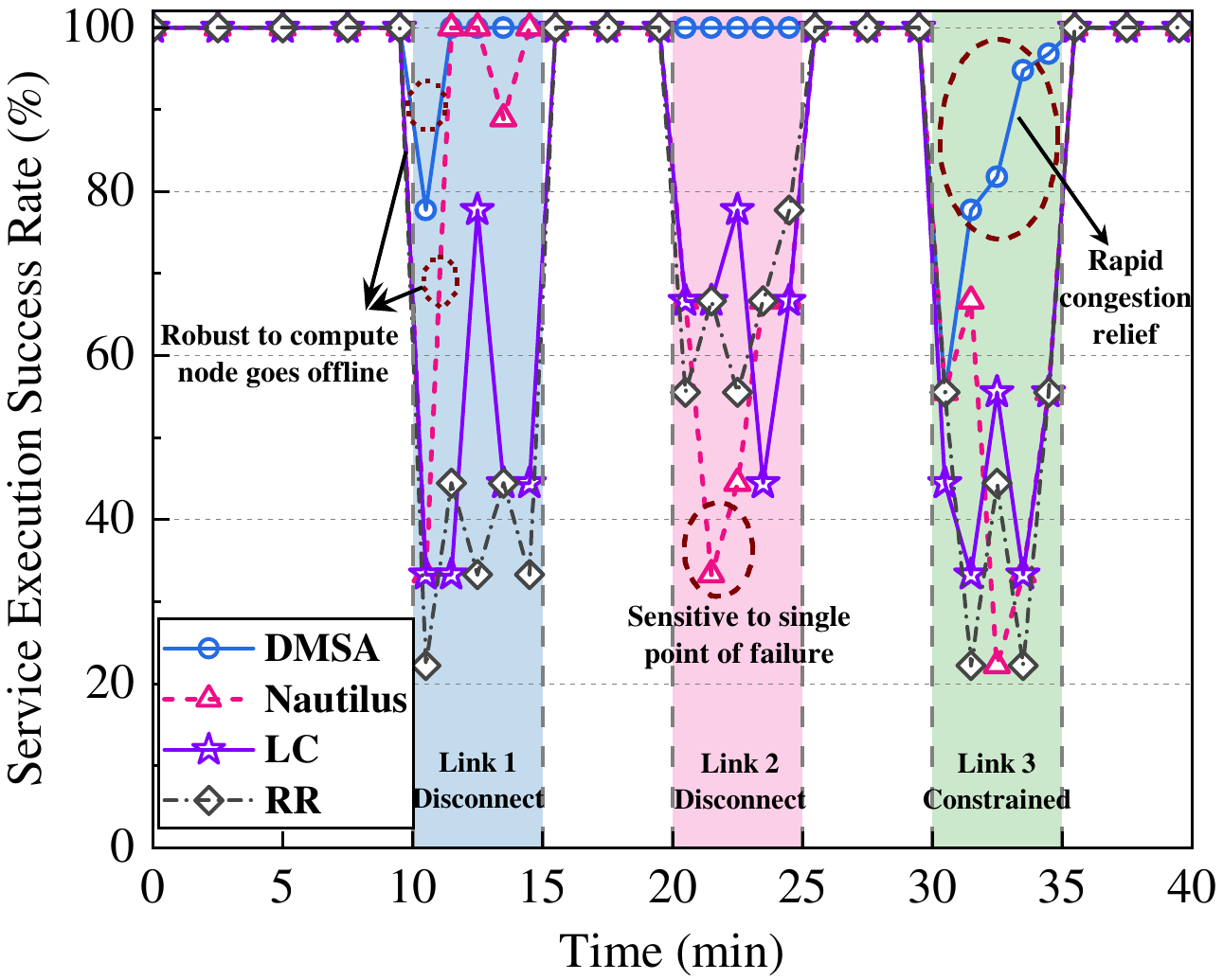}
            \caption{\small Low Load}
        \end{subfigure}
        \hfill
        \begin{subfigure}{0.9\columnwidth}
            \centering
            \includegraphics[width=\linewidth]{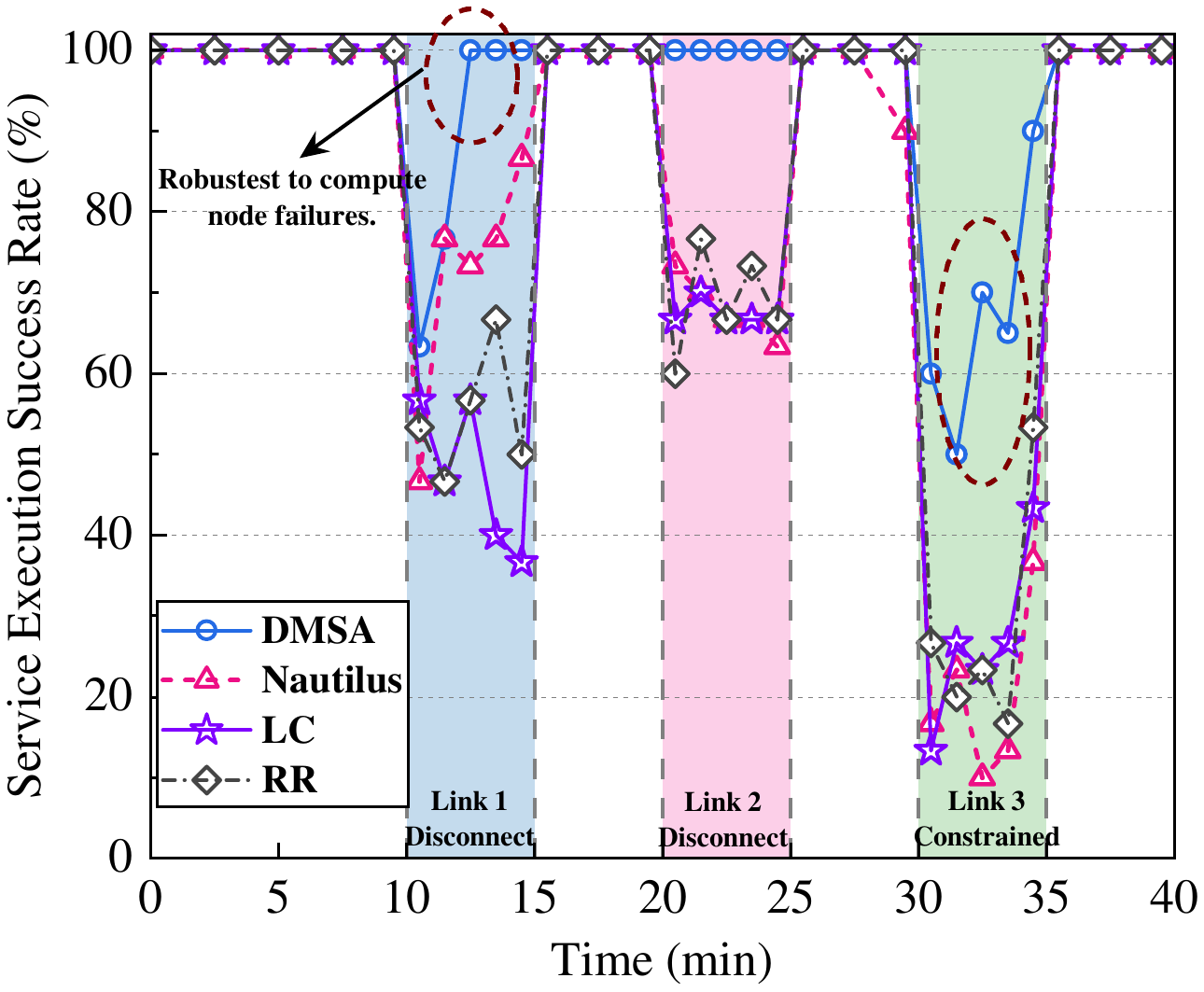}
            \caption{\small High Load}
        \end{subfigure}
        \vspace{-0.3em}
        \caption{\small The dynamic variation of service execution success rate of file download services with running time under (a) Low Load, and (b) High Load.}
        \vspace{0.3em}
        \label{fig17}
    \end{minipage}
    \vspace{-0.3em}
\end{figure}

\par Network event {\large \ding{193}} occurs in $20 \sim 25$ minutes, severing Link 2 between Switch 1 and Master node. In this case, the Master node becomes non-functional, rendering its instances inaccessible. Since there are fewer instances on the Master node than on Node 3, both LC and RR algorithms perform better during this network event than during {\large \ding{192}}. However, as a centralized MSA, Nautilus loses its network and node awareness upon the Master node's failure, rendering its scheduling paralysis, resulting in paramount response delay and diminished execution success rate. In contrast, the scheduling performance of DMSA is unaffected by the Master node's downtime due to its decentralized nature. The MAs on other nodes swiftly sense the non-responsiveness of the Master node through timeout, PSMS, and link bandwidth measurement mechanisms, thereby demoting the scheduling priority of instances on the Master node. This prompt action ensures response delays are briefly elevated before quickly returning to normal levels.

\par On the other hand, network event {\large \ding{194}} occurs between $30 \sim 35$ minutes, causing Link 3's bandwidth to fluctuate and plummet to $100$ Mbps. The LC and RR algorithms are powerless due to a lack of monitoring and sensing capabilities. User requests are continuously forwarded to the bandwidth-limited Link 3. Nautilus offers extremely limited performance improvements in complex edge network topologies since its scheduling relies on the nodes' and adjacent nodes' bandwidth loads. Conversely, DMSA demonstrates remarkable robustness and adaptability, quickly recovering from network fluctuations. If local-target node communication traverses Link 3, the monitoring module quickly detects the degradation in transmission rates leveraging active and passive measurement mechanisms. It reduces the target node's weight and recalibrates scheduling strategies, ensuring edge network performance remains significantly impervious to network fluctuations compared to other benchmarks.

% \vspace{-0.3em}

\subsubsection{Robustness Analysis of DMSA under Different Load Conditions}

\par Fig. \ref{fig16} shows the average execution success rates under various load conditions. DMSA improves execution success rates by approximately $10\% \sim 15\%$ compared to Nautilus, LC, and RR. Fig. \ref{fig16} (a)-(c) show that the advantages of DMSA in microservice scheduling become more pronounced as data volume and service arrival rates increase. For graphic-text services, DMSA maintains a stable execution success rate switching from low load to high load, while Nautilus, LC, and RR drop by $4.02\%$, $4.83\%$, and $4.52\%$, respectively. For video services, despite increasing loads, DMSA's success rate remains above $95\%$, compared to Nautilus's $3.18\%$ drop and LC and RR falling below $85\%$ under higher loads. In large file downloads, DMSA consistently exceeds a $90\%$ success rate, with Nautilus decreasing by $5.67\%$, and LC and RR hovering around $80\%$. Fig. \ref{fig17} highlights the dynamic changes in execution success rates for file download services. Under low loads, post-Link 1 failure, Nautilus quickly recovers performance upon detecting network changes. However, Nautilus's detection capability diminishes under high loads. In contrast, DMSA far outperforms the other baseline schemes.

\par
\begin{figure}[h]
    \centering
    \begin{minipage}[b]{0.5\textwidth}
        \centering
        \begin{subfigure}{0.9\columnwidth}
            \centering
            \includegraphics[width=\linewidth]{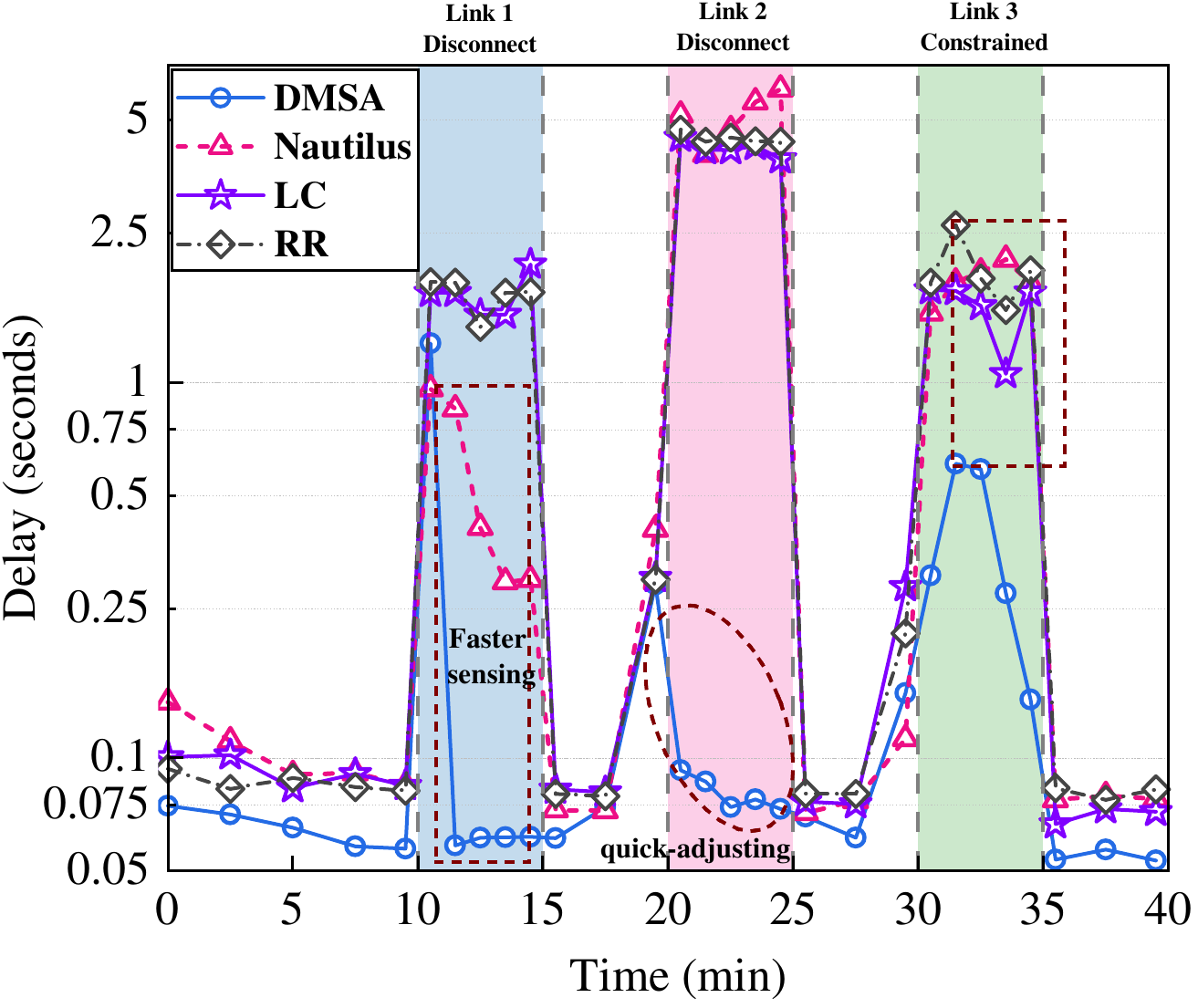}
            \caption{Response Delay}
        \end{subfigure}
        \hfill
        \begin{subfigure}{0.9\columnwidth}
            \centering
            \includegraphics[width=\linewidth]{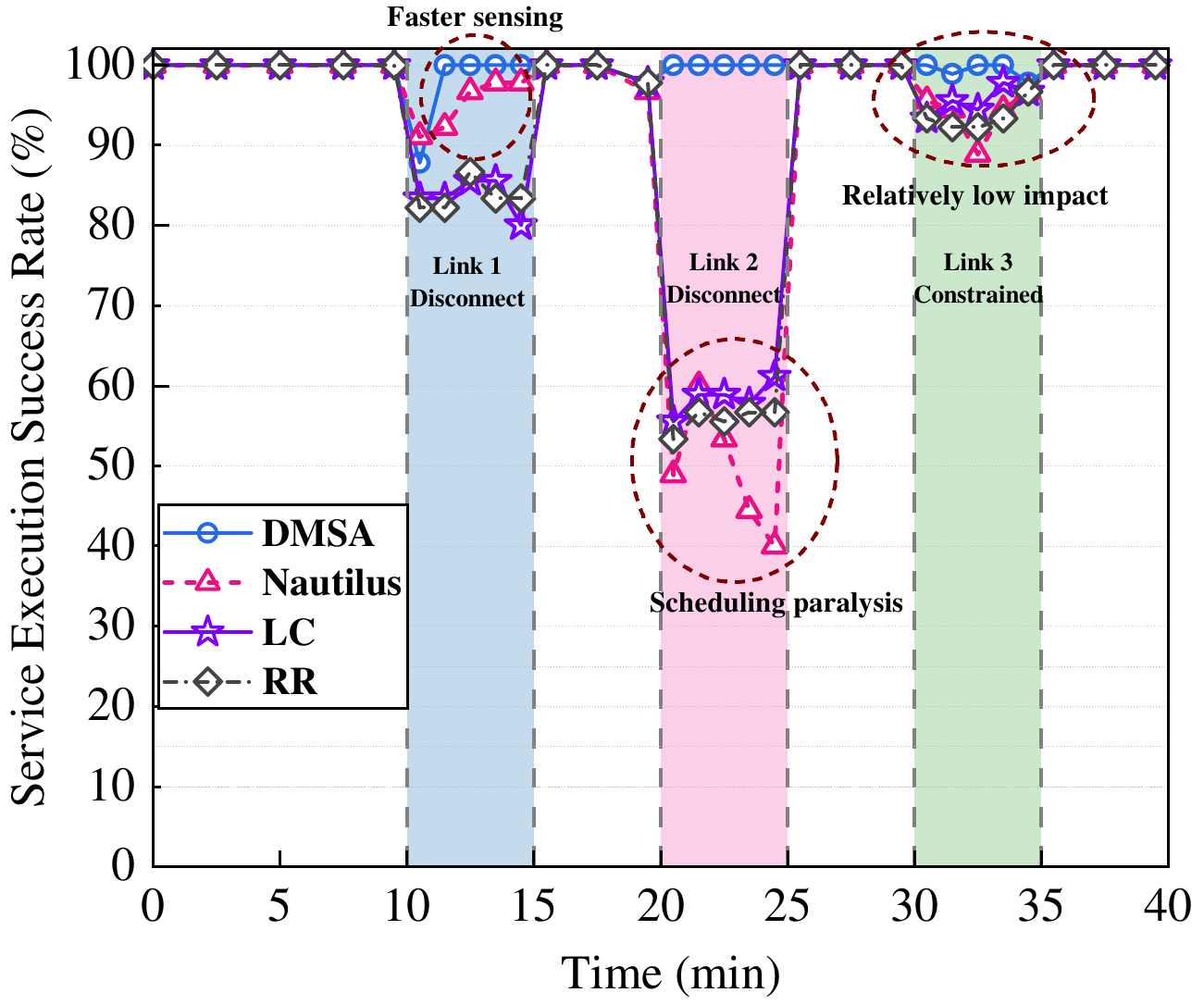}
            \caption{Execution Success Rate}
        \end{subfigure}
        \vspace{-0.3em}
        \caption{\small The dynamic variation of (a) response delay and (b) service execution success rate of graphic-text services with running time under Low Load.}
        \vspace{0.3em}
        \label{fig18}
    \end{minipage}
    \vspace{-0.3em}
\end{figure}

 Given the graphic-text service's small data size and low response delay requirements, Fig. \ref{fig18} shows the dynamics of response delay and execution success rate under low load. Upon Link 1 failure, although both DMSA and Nautilus sensed the change, DMSA outperforms Nautilus in quickly adjusting its scheduling strategies. When Link 2 disconnects, Nautilus and the unsusceptible LC and RR algorithms paralyze. Fig. \ref{fig18} (a) suggests that DMSA quickly generates new scheduling strategies based on node status changes, attributed to the efficient collaboration of its discovery, monitoring, and scheduling modules. Fig. \ref{fig18} (b) reveals that DMSA is largely unaffected by network fluctuations when Link 3 is constrained. Nonetheless, the response delay of graphic-text services is highly sensitive to network fluctuations. The response delay of Nautilus, LC, and RR increases dramatically, whereas DMSA presents resilience against this adverse effect. The discovery and monitoring modules identify the bandwidth constraints and prompt the scheduling module to adapt swiftly.

\section{Conclusion}
\par This paper proposes for the first time an innovative microservice architecture called DMSA, which breaks the traditional centralized microservice paradigm. DMSA delegates scheduling functions from the control plane to edge nodes. Moreover, it redesigns and implements the most core microservice discovery, monitoring, and scheduling modules for edge networks to achieve precise awareness of instance deployments, low monitoring overhead and measurement errors, and accurate dynamic scheduling, respectively. This empowers DMSA with robust resilience against various network events in edge networks with dispersed node locations and complex topologies, as well as intricate dynamic microservice dependencies. Finally, we implemente a physical verification platform for DMSA. Extensive empirical results demonstrate that compared with the state-of-the-art scheduling schemes, DMSA improves the service response delay and execution success rate by approximately $60\% \sim 75\%$ and $10\%\sim15\%$, respectively, and can effectively counteract link failures and network fluctuations, exhibiting superior robustness and adaptability.

\par As revealed in Fig. \ref{fig2}, effective microservice deployment significantly improves communication delay and load balancing. In future work, the optimization schemes for microservice deployment will be thoroughly investigated and implemented based on the proposed DMSA platform.

\footnotesize
\bibliographystyle{IEEEtran}
\bibliography{IEEEabrv,ref}
%\bibliographystyle{ieeetr}
%\bibliography{ref}
\end{document}